\documentclass[a4paper,fleqn]{cas-sc}
\usepackage[numbers,sort&compress]{natbib}
\usepackage{graphicx,amsmath,amssymb,amsfonts,mathrsfs,leftidx}
\usepackage{arydshln,amsthm,ragged2e,enumerate,longtable,float}
\usepackage{hyperref,subfig,tabularx,makecell,url,hhline}
\usepackage{multirow,booktabs,array,bm,adjustbox,indentfirst}
\usepackage[table,usenames,dvipsnames]{xcolor}
\definecolor{aa}{RGB}{175,238,238}
\definecolor{bb}{RGB}{255,255,255}
\newcolumntype{P}[1]{>{\raggedright\arraybackslash\footnotesize}m{#1}}
\newcolumntype{A}[1]{>{\centering\arraybackslash\footnotesize}m{#1}}

\theoremstyle{plain}

\begin{document}
\let\WriteBookmarks\relax
\shorttitle{Evolving Intelligent Complex Systems via Intellicise Networks: Architecture, Technologies, and Pathways}
\shortauthors{Zhang et al.}

\title[mode=title]{Evolving Intelligent Complex Systems via Intellicise Networks: Architecture, Technologies, and Pathways}

\ExplSyntaxOn
\cs_new_eq:NN \cas_saved_int_to_alph:n \int_to_alph:n
\cs_set_eq:NN \int_to_alph:n \int_eval:n
\ExplSyntaxOff

\stmaffsetup{
    size   = \footnotesize,
    shape  = \itshape,
    weight = \mdseries,
    color  = black,
    align  = \raggedright
}


\tnotemark[1]

\tnotetext[1]{This work was supported in part by the National Key R\&D Program of China under Grant 2020YFB1806905; in part by the National Natural Science Foundation of China under Grant 62501066; and in part by the S\&T Program of Hebei under Grant 262X0405D.}


\author[1]{Ping Zhang$^\dagger$}

\author[1,2]{Rui Meng$^\dagger$}
\cormark[1]
\ead{buptmengrui@bupt.edu.cn}

\author[1,2]{Xiaodong Xu}
\cormark[1]
\ead{xuxiaodong@bupt.edu.cn}

\author[1]{Song Gao}

\author[1]{Zixuan Huang}

\author[1]{Yaheng Wang}

\author[3]{Yinqiu Liu}

\author[3]{Ruichen Zhang}

\author[1]{Yiming Liu}

\author[4]{Kaiwen Yu}

\author[5]{Yaping Sun}

\author[6]{Han Meng}

\author[7]{Haonan Tong}

\author[8]{Huishi Song}

\author[9]{Qianqian Yang}

\author[10]{Shuoyao Wang}

\author[11]{Lexi Xu}

\author[12]{Qinghe Du}

\author[13]{Geng Sun}

\author[14]{Jiawen Kang}

\author[4]{Gang Wu}

\author[15,16,17]{Yiqing Zhou}

\author[18]{Haixia Zhang}

\author[19]{Zesong Fei}

\author[20]{Aimin Hao}

\author[21,22]{Ming Li}
\cormark[1]
\ead{mli@uwaterloo.ca}


\nonumnote{
$^\dagger$Ping Zhang and Rui Meng contributed equally
to this work and should be considered
co-first authors.
}

\cortext[cor1]{
Co-corresponding authors: Rui Meng and Xiaodong Xu.
}


\affiliation[1]{
 organization={State Key Laboratory of Networking
 and Switching Technology, Beijing University
 of Posts and Telecommunications},
 city={Beijing},
 postcode={100876},
 country={China}
}

\affiliation[2]{
 organization={Satellite Internet Testing Center,
 Xiong'an Aerospace Information Research Institute},
 city={Xiong'an},
 postcode={070001},
 country={China}
}

\affiliation[3]{
 organization={College of Computing and Data Science,
 Nanyang Technological University},
 city={Singapore},
 postcode={639798},
 country={Singapore}
}

\affiliation[4]{
 organization={National Key Laboratory of Wireless
 Communications, University of Electronic Science
 and Technology of China},
 city={Chengdu},
 postcode={611731},
 country={China}
}

\affiliation[5]{
 organization={Department of Broadband Communication,
 Pengcheng Laboratory},
 city={Shenzhen},
 postcode={518000},
 country={China}
}

\affiliation[6]{
 organization={Institute of Network and IT Technology,
 China Mobile Research Institute},
 city={Beijing},
 postcode={100053},
 country={China}
}

\affiliation[7]{
 organization={Key Laboratory of Target Cognition
 and Application Technology (TCAT),
 Aerospace Information Research Institute,
 Chinese Academy of Sciences},
 city={Beijing},
 postcode={100190},
 country={China}
}

\affiliation[8]{
 organization={ZGC Institute of Ubiquitous-X
 Innovation and Applications},
 city={Beijing},
 postcode={100083},
 country={China}
}

\affiliation[9]{
 organization={College of Information Science
 and Electronic Engineering, Zhejiang University},
 city={Hangzhou},
 postcode={310027},
 country={China}
}

\affiliation[10]{
 organization={College of Electronic and Information
 Engineering, Shenzhen University},
 city={Shenzhen},
 postcode={518000},
 country={China}
}

\affiliation[11]{
 organization={Research Institute,
 China United Network Communications Corporation},
 city={Beijing},
 postcode={100048},
 country={China}
}

\affiliation[12]{
 organization={School of Information and
 Communication Engineering,
 Xi'an Jiaotong University},
 city={Xi'an},
 postcode={710049},
 country={China}
}

\affiliation[13]{
 organization={College of Computer Science
 and Technology, Jilin University},
 city={Changchun},
 postcode={130012},
 country={China}
}

\affiliation[14]{
 organization={School of Automation,
 Guangdong University of Technology},
 city={Guangzhou},
 postcode={510006},
 country={China}
}

\affiliation[15]{
 organization={State Key Lab of Processors,
 Institute of Computing Technology,
 Chinese Academy of Sciences},
 city={Beijing},
 postcode={100190},
 country={China}
}

\affiliation[16]{
 organization={Beijing Key Laboratory of Mobile
 Computing and Pervasive Device},
 city={Beijing},
 postcode={100190},
 country={China}
}

\affiliation[17]{
 organization={University of Chinese
 Academy of Sciences},
 city={Beijing},
 postcode={100049},
 country={China}
}

\affiliation[18]{
 organization={Institute of Intelligent Communication
 Technology and Shandong Key Laboratory of
 Intelligent Communication and Sensing Computing
 Integration, Shandong University},
 city={Jinan},
 postcode={250061},
 country={China}
}

\affiliation[19]{
 organization={School of Information and Electronics,
 Beijing Institute of Technology},
 city={Beijing},
 postcode={100081},
 country={China}
}

\affiliation[20]{
 organization={State Key Laboratory of Virtual Reality
 Technology and Systems, Beihang University},
 city={Beijing},
 postcode={100191},
 country={China}
}

\affiliation[21]{
 organization={David R. Cheriton School
 of Computer Science, University of Waterloo},
 city={Waterloo},
 postcode={N2L 3G1},
 country={Canada}
}

\affiliation[22]{
 organization={Central China Institute
 of Artificial Intelligence},
 city={Zhengzhou},
 postcode={450046},
 country={China}
}

\begin{abstract}
Future engineering infrastructures are evolving into large-scale, open, heterogeneous, and wirelessly interconnected complex systems. These systems present significant challenges in optimizing network resource utilization, managing high-dimensional information spaces, and accommodating diverse business requirements. \textit{Intellicise (intelligent and concise)} networks, characterized by Intent-driven operation, semantic-native capability, and distributed intelligence, offer a promising paradigm for enabling such intelligent complex systems. This paper provides a systematic exploration of future intelligent complex systems from the perspective of intellicise networks. Specifically, we propose a cross-domain intelligent communication network architecture based on intellicise networks, grounded in information theory, system theory, game theory, and cybernetics. The architecture comprises a cross-layer organizational framework, multi-functional planes, and novel information flows. The cross-layer framework defines the vertical evolution from perception and cognition to decision, while the control, user, data, computation, intelligence, and security planes deliver horizontal intellicise capabilities. Moreover, data, knowledge, model, and task flows interconnect the various layers and planes, forming a closed-loop process that derives simplicity from high-level intelligene while concurrently pursuing enhanced. Building on this architecture, we review key enabling technologies, tracing their evolution from semantic extraction to intent understanding, from heterogeneous resource integration to self-configuration and self-optimization, from generative artificial intelligence (AI) to agentic AI, and from embodied AI to symbodied AI. Additionally, we present a case study on intellicise networks for embodied agent communications and discuss representative applications and services for intelligent complex systems. Finally, we summarize key challenges and outline future research directions to guide the advancement of intelligent complex systems enhanced by intellicise networks.
\end{abstract}

\begin{keywords}
Intellicise (intelligent and concise) network, complex system, Artificial Intelligence (AI).
\end{keywords}

\maketitle

\ExplSyntaxOn
\cs_set_eq:NN \int_to_alph:n \cas_saved_int_to_alph:n
\ExplSyntaxOff

\section{Introduction}
Engineering technology has become a fundamental driving force for human society, and large-scale engineering systems are increasingly regarded as strategic infrastructures for national development and industrial upgrading. In the coming stage of technological evolution, engineering and information systems are evolving from isolated technical facilities into large-scale, open and heterogeneous systems \cite{cao2025advancing,feng2025towards}. Major engineering scenarios, such as space-air-ground-sea integrated networks (SAGSINs), industrial Internet of Things (IIoT), and so on, are no longer supported by separate functions \cite{meng2026semantic}. Instead, they evolve with changing demands and continuously interact with external environments \cite{11417150}. Therefore, future engineering infrastructures should not be viewed merely as simple communication networks, they should be modeled as complex systems involving diverse forms of interaction.

Complex systems are characterized by a large number of interacting components, which causes a nonlinear coupling relationships and uncertainty behaviors \cite{yang2023complex}. In future infrastructures, such complexity could be highlighted because of the increasingly connected entities including humans, machines, objects and intelligence agents within the same networking system \cite{meng2026secure}. 
As a result, traditional separated optimization of communication, computation, intelligence, and control is insufficient to support the efficient operation and autonomous evolution of future intelligent complex systems.
First, the number of network nodes is increasing rapidly. The frequent and dynamic interactions make centralized coordination and predefined network configurations difficult to scale. A large number of interacted entities are connected in real time, which makes the information no longer propagates along simple and predictable paths. Even local changes may be amplified through interactions and lead to nonlinear system-level effects \cite{qiu2025disentangled}. 
Second, the amount of data generated within the system is becoming massive. Multi-modal sensing data, environmental states, and task feedback are continuously produced, making it difficult to identify important information. 
Third, the system is becoming increasingly diverse in terms of service requirements, which is hard to design an unified and statical modeling strategy to fit different objectives in various domains.

These challenges indicate that future complex systems require a new paradigm that transition from traditional communication systems toward intelligent, semantic-aware, and task-oriented network architectures \cite{you2021towards}. Intellicise networks provide a promising solution to this problem \cite{zhang2024intellicise,zhang2025modernSemantic6GIntellicise}. Different from traditional communication networks that mainly pursue  bit-level accuracy, intellicise networks focus on semantic-level transmission, which achieves by extracting task-relevant semantics to simplify information interaction and enable autonomous network organization \cite{nan2023physical}. By introducing endogenous intelligence into information transmission, resource allocation and network management \cite{peng2025activebatteryfreerydbergatomic}, intellicise networks enable an intelligent complex systems that can continuously sense the environment, transmit information, processes knowledge and make decisions, which improves the communication efficiency and enhance the adaptability of complex systems under dynamic environment.


Although existing studies have investigated complex systems from different perspectives\cite{krakauer2026large, lu2022survey, tang2021state}, a systematic framework that explains how intellicise networks enable future intelligent complex systems remains nascent.
It is necessary to explain how intellicise networks can transform complex systems from static infrastructures into intelligence systems to realize autonomous evolution. Motivated by this gap, this article provides a systematic perspective on future intelligent complex systems from the viewpoint of intellicise networks. The main contributions are summarized as follows.
\begin{itemize}
\item We propose a cross-domain intelligent communication network architecture based on intellicise networks for enabling future intelligent complex systems. The architecture integrates a cross-layer organizational structure, multi-functional planes, and novel information flow structures for perception, transmission, computing, and decision-making. 

\item We provide key enabling technologies for the evolution of intelligent complex systems and summarize representative applications and services enabled by intellicise networks.

\item We present a case study to verify that the intellicise networks can improve the performance of embodied agent communications.

\item We discuss future research directions for intelligent complex systems enabled by intellicise networks.
\end{itemize}

\begin{figure}
\centering
\includegraphics[width=\linewidth,height=.58\textheight,keepaspectratio]{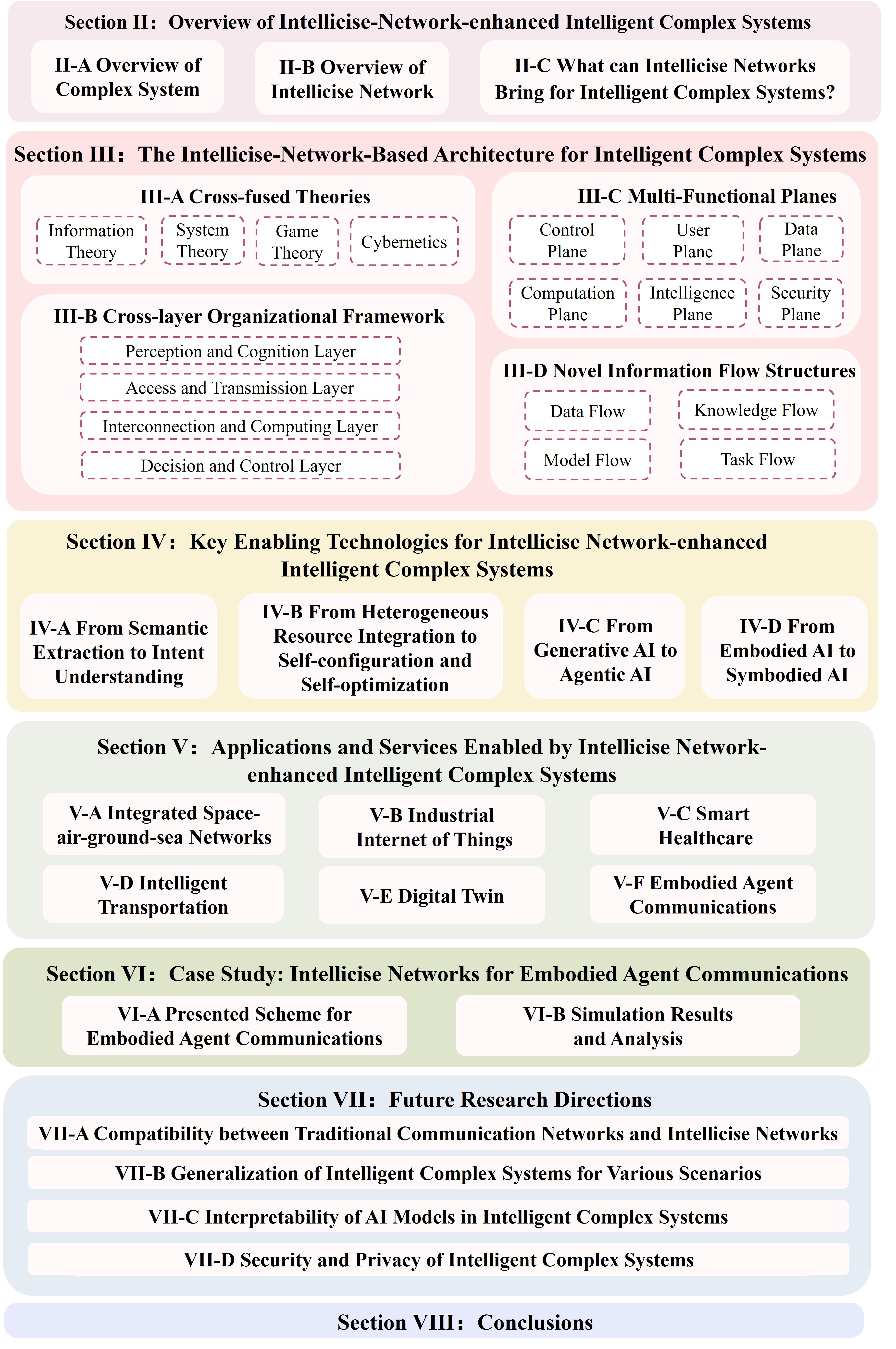}
\caption{The outline of this paper.}
\label{fig_outline}
\end{figure}

As shown in Fig. \ref{fig_outline}, the rest of this paper is organized as follows. Section \ref{section2} provides an overview of complex systems, intellicise networks, and their interactions in future intelligent complex systems. Section \ref{section3} introduces the proposed cross-domain intelligent communication network architecture. Section \ref{section4} reviews key enabling technologies for intelligent complex systems. Section \ref{section5} presents applications and services supported by intelligent complex systems. Section \ref{section6} presents a case study to verify the effectiveness of intellicise networks in enhancing embodied agent communications. Section \ref{section7} outlines future research directions. Finally, Section \ref{section8} concludes this paper.

\section{Overview of Intellicise-Network-enhanced Intelligent Complex Systems}
\label{section2}
\begin{figure}
\centering
\includegraphics[width=\linewidth,height=.58\textheight,keepaspectratio]{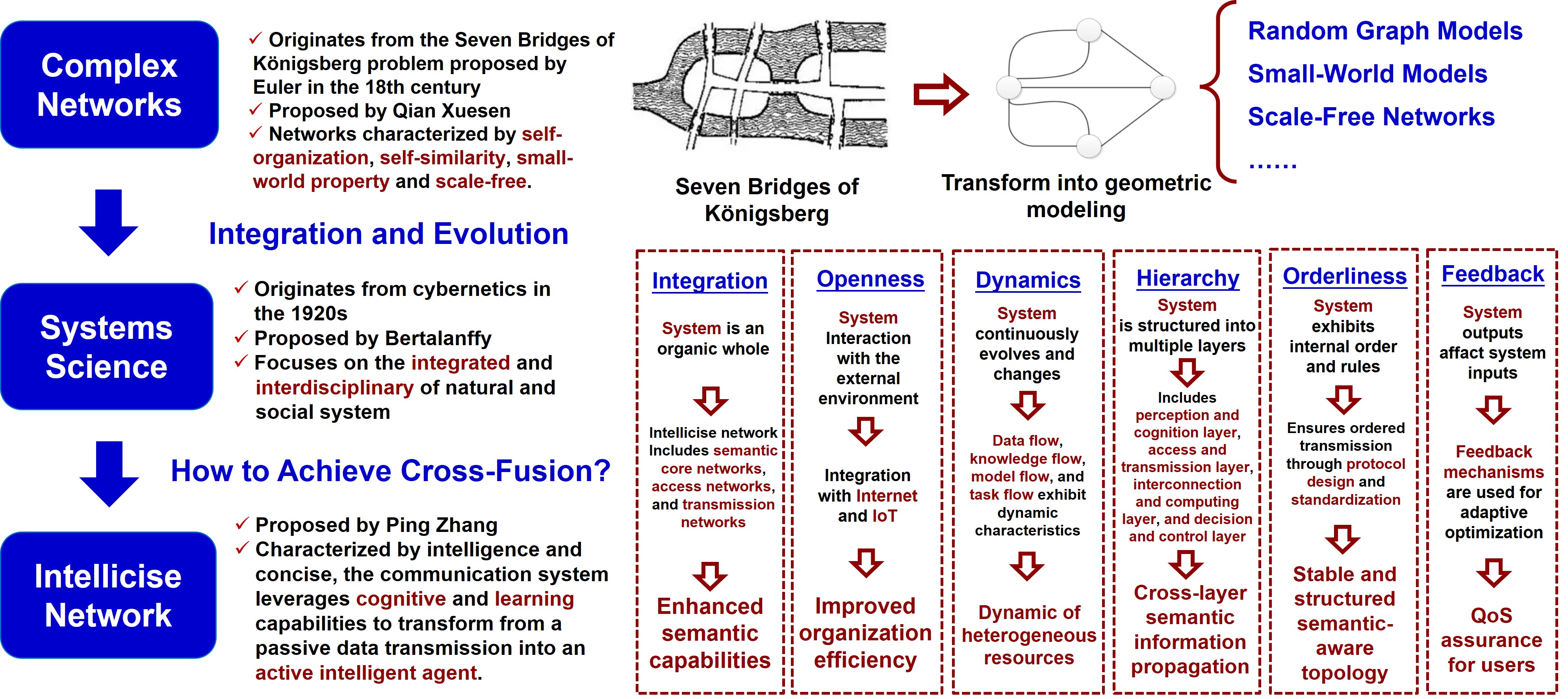}
\caption{The overview of system science-based intellicise networks, where complex networks provide interaction modeling tools, systems science offers fundamental principles for complex system management, and intellicise networks inherit these principles to support semantic enhancement, resource organization, cross-layer propagation, topology stability, and service assurance.}
\label{figcomplex_system}
\end{figure}

\subsection{Overview of Complex System}

\subsubsection{What is Complex System?}
A complex system consists of a large number of interacting components or entities, often with direct or indirect nonlinear relationships among them. These interactions give rise to overall system behaviors and characteristics that individual components do not possess\cite{newman2011complex}. Complex systems typically exhibit properties such as self-organization, uncertainty, and dynamic evolution. They are widely found in both natural systems and human society, such as the Internet, economic systems, and social structures\cite{yang2023complex}.


\subsubsection{What are the Challenges Faced by Complex System?}

Complex systems face three major challenges as follows.
\begin{itemize}
\item 
\textbf{Nonlinear Information Dissemination:} In complex systems, differences in information dissemination methods can produce nonlinear effects, which may lead to explosive information growth and temporal misalignment in propagation\cite{ding2024turing}. These effects increase the difficulty of network control and optimal resource utilization.
\item 
\textbf{High Dimensionality of the Information Space:} Extensive interactions of complex information within complex systems cause the information space to expand rapidly, potentially leading to the curse of dimensionality\cite{han2024semantic}. This exacerbates the difficulty of representing the information-bearing space.
\item 
\textbf{High Complexity of Information Processing:} Complex systems typically contain a large number of nodes, each carrying varying business requirements, and the coupling relationships among these nodes are intricate\cite{hu2024exploiting}. These factors collectively contribute to the high complexity of information processing tasks.
\end{itemize}

\subsection{Overview of Intellicise Networks}
The universal framework of \textit{Intellicise (intelligent and concise)} networks includes the brain for intellicise networks, intellicise signal processing, intellicise information transmission, intellicise network organization, and intellicise service bearing\cite{zhang2024intellicise}.
By integrating theories such as information theory, complex science, and system theory, intellicise networks can not only achieve information transmission, but also autonomously optimize network organization based on service demands. Unlike classical communication systems that aim to transmit signals accurately at the bit level, the intellicise networks extract the semantics directly related to the task for transmission. By reducing redundant information and simplifying management processes, it leverages high-level intelligence to simplify information interaction and network operation, while improving system efficiency, adaptability, and service performance, which achieves lower complexity, higher network efficiency and stronger adaptability, thus providing support for intelligent complex systems.

\subsection{What can Intellicise Networks Bring for Intelligent Complex Systems?}
Systems science, which integrates complex network theory with other disciplines, provides a fundamental methodological pathway for understanding and managing complex systems. 
As illustrated in Fig. \ref{figcomplex_system}, the intellicise networks interact closely with systems science, offering the following significant insights for the development of intelligent complex systems.
\begin{itemize}
\item 
\textbf{Semantic Native Capability Enhancement:} Semantic communication, as a key technology for intellicise networks, can greatly improve transmission efficiency and reliability\cite{zhang2022toward}. Drawing on the holistic principle, which states that a system is an organic whole, systems science can further enhance the semantic native capabilities of intellicise networks.
\item 
\textbf{Intellicise Networking Performance Improvement:} To update the semantic knowledge base (KB), intellicise networks need to be open and integrated with the Internet, IoT, and other networks \cite{sun2023semanticKB}. According to the openness principle, which holds that a system must communicate with the external environment to sustain development, systems science can boost the performance of intellicise networking.
\item 
\textbf{Heterogeneous Resource Dynamic Management:} The demands for communication resources, computing resources, intelligent resources, perception resources, and others in intellicise networks are constantly changing, requiring dynamic management\cite{zhang2024intellicise}. Based on the dynamic principle, which recognizes that systems are in continuous motion, change, and development, systems science can improve the dynamic management of heterogeneous resources in intellicise networks.
\item 
\textbf{Cross-Layer Propagation of Semantic Information:} The cross-layer semantic interaction in intellicise networks can enhance network performance\cite{zhang2025modernSemantic6GIntellicise}. Following the hierarchical principle, which divides systems into different levels, systems science can strengthen the ability of semantic information to propagate across layers.
\item 
\textbf{Stability and Orderliness of Intellicise Topology:} Intellicise networks adhere to protocols and standards to ensure orderly communication among devices and the secure operation of the network\cite{zhang2026towardss}. In line with the orderliness principle, which asserts that order and rules exist within a system, systems science can guarantee the stability and orderliness of intellicise topology.
\item 
\textbf{User Service Quality Assurance:} Intellicise networks perform optimization and upgrades based on user experience feedback\cite{zhang2022toward}. Applying the feedback principle, where the output of a system influences its input, systems science can ensure the quality of service for users.
\end{itemize}

\section{The Intellicise-Network-based Architecture for Intelligent Complex Systems}

\label{section3}
\begin{figure}
\centering
\includegraphics[width=\linewidth,height=.58\textheight,keepaspectratio]{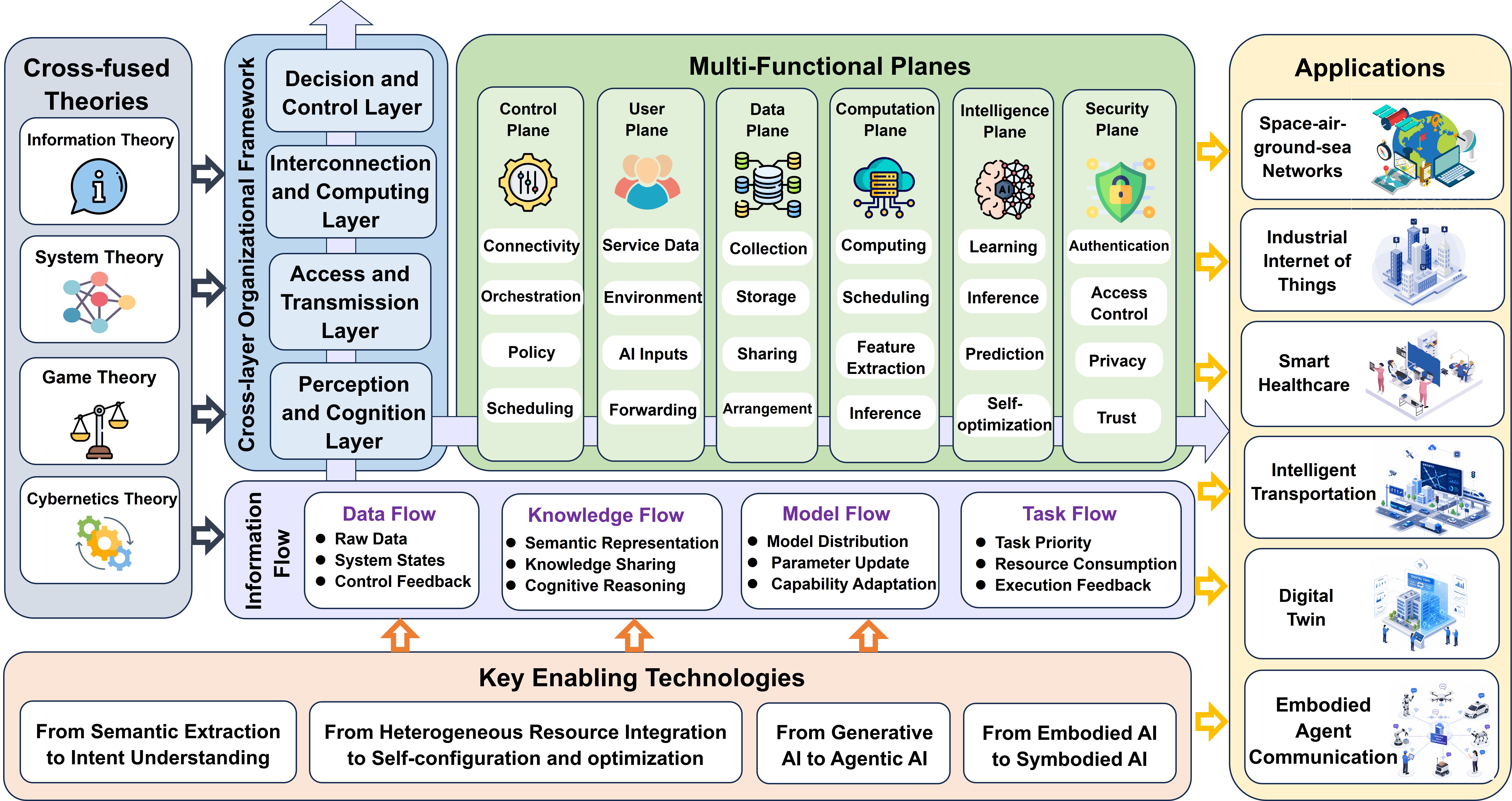}
\caption{The proposed cross-domain intelligent communication network architecture based on intellicise networks, where cross-fused theories provide the theoretical foundations, the cross-layer organizational framework defines the vertical evolution from perception and cognition to decision and control, and the multi-functional planes offer horizontal capabilities for control, service delivery, data management, computation, intelligence, and security. The data, knowledge, model, and task flows connect different layers and planes to form a closed-loop process, while key enabling technologies support the practical implementation of the framework and bridge it with representative applications and services.}
\label{figframework}
\end{figure}

To leverage intellicise networks for improving intelligent complex systems, we propose a cross-domain intelligent communication network architecture, as shown in Fig. \ref{figframework}. Specifically, cross-fused theories provide the fundamental principles for understanding and controlling complex system behaviors, including information representation, system evolution, strategic interaction, and feedback control. Based on these principles, the cross-layer framework defines the vertical structure of the system, while the multi-functional planes provide horizontal capabilities for control, data, computation, intelligence, security, and service delivery. The novel information flow structures further describe how data, knowledge, models, and tasks circulate across layers and planes, forming a closed-loop process from perception and transmission to computing, decision-making, and feedback. Key enabling technologies serve as the practical support for enhancing the performance of intelligent complex systems, and finally connect the architectural design with diverse applications and services. In the following subsections, we first explain how the cross-fused theories inform the design of the architecture. Then, we describe the cross-layer organizational framework that defines its vertical structure. Furthermore, we explain how the multi-functional planes provide capabilities across the layers. Finally, we show how the information flows connect the layers and planes to support closed-loop operation as follows.

\subsection{Cross-fused Theories}

\begin{table}
	\centering
	\caption{Summary of Cross-Fused Theories and Their Intersection with Intellicise Networks}
	\label{tab:cross_fused_theories}
	\renewcommand{\arraystretch}{1.4}
	\setlength{\tabcolsep}{5pt}
	
	\providecommand{\gc}{\cellcolor{gray!15}}
	
	\begin{adjustbox}{max width=\linewidth}
\begin{tabular}{
		|>{\raggedright\arraybackslash}m{2.7cm}
		|>{\raggedright\arraybackslash}m{4.0cm}
		|>{\raggedright\arraybackslash}m{4.5cm}
		|>{\raggedright\arraybackslash}m{5.4cm}|
	}
		\hline
		\textbf{Theory}
		& \textbf{Foundation}
		& \textbf{Key Theories}
		& \textbf{Intersection with Intellicise Networks}
		\\ \hline
		
		\gc Information Theory
		&
		\gc Shannon's mathematical foundation for modern communication
		&
		\gc $\bullet$ Statistical fidelity of transmitted symbols.\par
		$\bullet$ Semantic information theory.
		&
		\gc Drives the evolution from bit-level transmission to semantic-level understanding.
		\\ \hline
		
		System Theory
		&
		An organic whole with emergent capabilities arising from interactions among microscopic components
		&
		 $\bullet$ Complex adaptive systems theory.\par
		$\bullet$ Synergetics.\par
        $\bullet$ System of systems engineering.
		&
		Addresses challenges involving strong nonlinearity, multi-node interactions, and adaptive evolution.
		\\ \hline
		
		\gc Game Theory
		&
		\gc Rational decision-makers interacting strategically to maximize individual or collective payoffs or utilities
		&
		\gc $\bullet$ Non-cooperative game.\par
		$\bullet$ Markov game.\par
        $\bullet$ Evolutionary game.
		&
		\gc Enables cognitive radio and dynamic spectrum sharing.
		\\ \hline
		
		Cybernetics
		&
		Feedback-based control and communication in machines, organisms, and composite systems.
		&
		 $\bullet$ Engineering cybernetics.\par
		$\bullet$ Black-box theory.\par
        $\bullet$ Nonlinear control theory.
		&
		Bridges closed-loop control between the physical world and digital intelligence.
		\\ \hline
		
	\end{tabular}
\end{adjustbox}
\end{table}


The design of the cross-domain intelligent communication network architecture must be anchored in a robust theoretical framework that reflects the characteristics of intelligent complex systems. To manage massive, heterogeneous data, information theory must transcend bit-level accuracy to characterize semantic value and task-specific relevance. For large-scale interactions among network entities, systems theory is essential to describe interdependencies, collective behaviors, and evolutionary processes. Given autonomous entities with disparate objectives, game theory becomes crucial for coordinating cooperation among multiple decision-makers. Moreover, dynamic environments and fluctuating task requirements necessitate cybernetics to enable feedback control and adaptive closed-loop operations. Collectively, these theoretical foundations underpin the proposed architecture and drive the evolution of intelligent complex systems, as summarized in Table \ref{tab:cross_fused_theories}.

\subsubsection{Information Theory}

Shannon information theory lays the mathematical foundation for modern communication theory \cite{shannon1948mathematical}. The theory introduces the concept of information entropy to provide a quantitative measure of information, and it describes a communication process in which an information source emits a message, an encoder converts the message into a form suitable for transmission over the channel, the signal propagates through a noisy channel, and a decoder reconstructs the message, which is then received at the destination. Throughout this process, the statistical fidelity of transmitted symbols is of central concern and has been continuously improved through the technological evolution from 1G to 5G. In recent years, the rapid advancement of artificial intelligence (AI) and its gradual integration with information theory have given rise to new theories such as semantic information theory, which formulates concepts like synonymous mapping, semantic entropy, and semantic mutual information and extends the Shannon framework into the dimension of semantics \cite{zhang2026beyond}.

\subsubsection{System Theory}
General systems theory posits that a complex system is not merely a mechanical aggregate of isolated physical components; rather, it constitutes an organic whole in which entirely new capabilities emerge at the macroscopic level through intricate interactions among microscopic constituents \cite{von1968general}. Over time, system theory has evolved into several derivative theories, including complex adaptive systems \cite{holland1995hidden}, synergetics \cite{haken1983synergetics}, and system of systems engineering \cite{jamshidi2008system}. These theories provide systematic support for understanding how local interactions, feedback relationships, and structural evolution collectively shape system-level behaviors.
System theory is now extensively applied to address challenges in complex systems characterized by strong nonlinearity, multi-node interactions, and adaptive evolution, such as agentic AI \cite{miehling2025agentic} and smart grids \cite{ohanu2024comprehensive}.

\subsubsection{Game Theory}


Game theory studies how multiple rational decision-makers act in strategic interactions, where their behaviors mutually influence and constrain one another to maximize their own payoffs or utilities \cite{vonneumann1947theory}. Its applications have continuously expanded into a wide range of fields, as exemplified by multi-agent reinforcement learning (MARL) based on Markov games \cite{littman1994markov} and generative adversarial networks (GANs) \cite{goodfellow2014generative}. By characterizing competition, cooperation, incentives, and equilibrium, game theory provides a principled foundation for decentralized coordination among network entities. In the communication  and network domain, game theory has also been widely applied to various technologies, including cognitive radio and dynamic spectrum sharing \cite{niyato2008competitive}, network incentive mechanisms and resource pricing in federated learning \cite{kang2019incentive}, as well as physical layer security and anti-jamming defense \cite{saad2009physical}.

\subsubsection{Cybernetics}

Cybernetics studies the fundamental laws of control and communication within and between machines, biological organisms, and composite systems. Norbert Wiener established the foundational framework of cybernetics by unifying the operational principles of biological and mechanical systems through information and feedback mechanisms \cite{wiener1948cybernetics}. Subsequently, H. S. Tsien (Xuesen Qian) extended these theoretical constructs into engineering applications by systematically formalizing Engineering Cybernetics \cite{tsien1954engineering}. Furthermore, W. Ross Ashby systematically elucidated black-box theory and the law of requisite variety \cite{ashby1956introduction}. Nonlinear control theory, based on the foundational stability criterion proposed by A. M. Lyapunov \cite{lyapunov1892general}, was further developed within the differential geometric control framework \cite{isidori1985nonlinear}, thus extending the applicability boundaries of cybernetics to complex nonlinear systems in the real physical world. Collectively, these theories provide the indispensable theoretical underpinnings for modern complex systems and automation technologies. Currently, the emergence of imitation learning \cite{zare2024survey}, Embodied AI \cite{duan2022survey}, and NeuroAI \cite{chiel2026brain} underscores the continuous evolution of cybernetics in the era of AI.

\subsection{Cross-layer Organizational Framework}

\begin{table}
	\centering
	\caption{Summary of the Cross-layer Organizational Framework for Intelligent Complex Systems}
	\label{tab:cross_layer_framework}
	\renewcommand{\arraystretch}{1.4}
	\setlength{\tabcolsep}{5pt}
	
	\providecommand{\gc}{\cellcolor{gray!15}}
	
	\begin{adjustbox}{max width=\linewidth}
\begin{tabular}{
		|>{\raggedright\arraybackslash}m{4.5cm}
		|>{\raggedright\arraybackslash}m{5.5cm}
		|>{\raggedright\arraybackslash}m{7cm}|
	}
		\hline
		\textbf{Layer}
		& \textbf{Core Role}
		& \textbf{Key Functions}
		\\ \hline
		
		\gc Decision and Control Layer
		&
		\gc Autonomous evolution and closed-loop decision-making
		&
		\gc $\bullet$ Intent understanding and task decomposition.\par
		$\bullet$ Formulation of optimal or near-optimal strategies.\par
		$\bullet$ System-wide execution orchestration.
		\\ \hline
		
		Interconnection and Computing Layer
		&
		Converged computing and distributed collaboration
		&
		$\bullet$ Unified orchestration of heterogeneous resources.\par
		$\bullet$ Edge/fog/cloud-based collaborative distributed computing.\par
		$\bullet$ Compute offloading and edge inference.
		\\ \hline
		
		\gc Access and Transmission Layer
		&
		\gc Semantic-aware reliable transmission
		&
		\gc $\bullet$ Heterogeneous multi-access.\par
		$\bullet$ Semantic-aware information transmission.\par
		$\bullet$ Dynamic modulation, coding, and resource allocation.
		\\ \hline
		
		Perception and Cognition Layer
		&
		Semantic modeling of the physical world
		&
		$\bullet$ Multi-modal data acquisition.\par
		$\bullet$ Semantic feature extraction.\par
		$\bullet$ Knowledge graph construction.
		\\ \hline
		
	\end{tabular}
\end{adjustbox}
\end{table}

Traditional communication networks follow a rigid vertical hierarchy with strict layer isolation, where each layer operates independently and interacts with adjacent layers through well-defined protocols. While this architecture offers clarity and modularity, it falls short in intelligent complex systems, where cross-layer interaction, dynamic resource coupling, and task-driven coordination are essential. To address this limitation, the proposed architecture organizes the capabilities of intellicise networks into four cascading layers \cite{zhang2026towards}, ranging from the perception and cognition layer at the bottom to the decision and control layer at the top, which are summarized in Table \ref{tab:cross_layer_framework}.


\subsubsection{Perception and Cognition Layer}

This layer serves as the foundational interface between the physical world and the intellicise networks, responsible for raw data acquisition and initial semantic interpretation. In intelligent complex systems, sensing modalities such as vision, radar, audio, and environmental sensors continuously capture physical information, including spatial geometry, motion states, and environmental occupancy \cite{letaief2019roadmap}. The perception and cognition layer then performs semantic extraction and knowledge graph construction \cite{fan2025kgrag}, transforming raw observations into semantic representations aligned with task objectives. The refined semantic information generated at this layer is forwarded upward to support resource orchestration and decision-making, while knowledge about the specific environment is fed back to maintain coherence of the shared semantic KB \cite{liu2025csgo}.

\subsubsection{Access and Transmission Layer}

This layer is responsible for heterogeneous multi-access and reliable transmission across different domains within intelligent complex systems. Building on traditional physical layer transmission, it evolves to support semantic-aware transmission, prioritizing information critical to the task over raw data fidelity \cite{gunduz2022beyond}. The access and transmission layer coordinates radio access technologies across terrestrial, aerial, and satellite segments, dynamically adapting modulation, coding, and resource allocation strategies based on channel states and task requirements \cite{cao2026s}. Within the cross-layer framework, this layer receives semantic representations from the perception and cognition layer and delivers them to the interconnection and computing layer with minimal delay and maximal fidelity, forming the backbone of the vertical information pipeline in intelligent complex systems.

\subsubsection{Interconnection and Computing Layer}

This layer manages the unified orchestration of heterogeneous resources and supports collaborative distributed computing across intelligent complex systems. It provides a cohesive abstraction over computing, communication, and storage resources distributed across edge, fog, and cloud nodes \cite{yuan20256g}. By integrating compute offloading, edge inference, and federated learning mechanisms, the interconnection and computing layer transforms transmitted semantic information into actionable intelligence. The intelligence plane and computation plane jointly enable dynamic scheduling of resources and model deployment at this layer \cite{xiao2026sanet}.

\subsubsection{Decision and Control Layer}

This layer occupies the apex of the cross-layer framework and is responsible for intent understanding, task decomposition, autonomous decision-making, and closed-loop control within intelligent complex systems. Drawing on semantic information, knowledge structures, and environmental states forwarded from lower layers, it formulates optimal or near-optimal strategies that govern system-wide resource allocation, task prioritization, and execution scheduling \cite{velasco2021end}. The decision and control layer closes the loop from perception to action by issuing directives downward through the interconnection and computing layer and the access and transmission layer, thereby realizing self-configuration, self-optimization, and adaptive evolution across the entire intelligent complex system.

\subsection{Multi-Functional Planes}

\begin{table}
	\centering
	\caption{Comparison of Multi-functional Planes in Traditional and Intellicise Networks}
	\label{tab:multi_functional_planes}
	\renewcommand{\arraystretch}{1.35}
	\setlength{\tabcolsep}{5pt}
	
	\providecommand{\gc}{\cellcolor{gray!15}}
	
	\begin{adjustbox}{max width=\linewidth}
\begin{tabular}{
		|>{\raggedright\arraybackslash}m{2.2cm}
		|>{\raggedright\arraybackslash}m{3.8cm}
		|>{\raggedright\arraybackslash}m{11cm}|
	}
		\hline
		\textbf{Plane}
		& \textbf{Core Role}
		& \textbf{Comparison with Traditional Networks}
		\\ \hline
		
		\gc Control Plane
		&
		\gc Central planning
		&
		\gc $\bullet$ \textbf{Traditional:} Focuses primarily on signaling-based network control.\par
		$\bullet$ \textbf{Intellicise:} Extends to cross-domain and cross-task collaborative management.
		\\ \hline
		
		User Plane
		&
		Service data delivery and environmental awareness
		&
		$\bullet$ \textbf{Traditional:} Supports packet forwarding, tunneling, and traffic management.\par
		$\bullet$ \textbf{Intellicise:} Further carries environmental sensing information and AI-task-related data.
		\\ \hline
		
		\gc Data Plane
		&
		\gc Foundational data-resource base
		&
		\gc $\bullet$ \textbf{Traditional:} Data management is closely coupled with the forwarding process.\par
		$\bullet$ \textbf{Intellicise:} Provides unified data collection, storage, processing, and sharing services for multiple planes.
		\\ \hline
		
		Computation Plane
		&
		Computing-power infrastructure
		&
		$\bullet$ \textbf{Traditional:} Lacks a dedicated mechanism for computing resource scheduling.\par
		$\bullet$ \textbf{Intellicise:} Enables unified management of distributed computing resources.
		\\ \hline
		
		\gc Intelligence Plane
		&
		\gc System-wide intelligence
		&
		\gc $\bullet$ \textbf{Traditional:} Does not provide native intelligence capabilities.\par
		$\bullet$ \textbf{Intellicise:} Supports full-lifecycle AI capabilities, including model training, deployment, inference, evaluation, and updating.
		\\ \hline
		
		Security Plane
		&
		System-wide security and trust
		&
		$\bullet$ \textbf{Traditional:} Focuses on link protection, identity authentication, and user-data security.\par
		$\bullet$ \textbf{Intellicise:} Extends protection to data, models, knowledge, tasks, and cross-plane collaborative processes.
		\\ \hline
		
	\end{tabular}
\end{adjustbox}
\end{table}

In the proposed architecture, the capabilities of intellicise networks are organized into six functional planes: the control plane for heterogeneous resource management, the user plane for service input and output, the data plane for multidimensional data storage and processing, the computation plane for computing power management, the intelligence plane for AI capabilities, and the security plane for ensuring the information security and stability of all planes, share information and collaborate with one another within a closed-loop intelligent control system. These six planes link the environment, knowledge, task intent, and network control, thereby enabling intelligent complex systems to process information more reliably and efficiently. The core role and the comparison with traditional networks of these six planes are summarized in Table \ref{tab:multi_functional_planes}.

\subsubsection{Control Plane}

The control plane serves as the central planning element in intelligent complex systems, responsible for the coordinated management of connectivity, resources, policies, and task execution. In traditional networks, the control plane relies primarily on signaling to perform access control, authentication and authorization, mobility management, session management, policy control, and radio resource management, thereby ensuring network connectivity and service availability. In contrast, within intelligent complex systems, the control plane enables coordination across different planes, domains, and tasks \cite{velasco2021end}. 


\subsubsection{User Plane}

The user plane is responsible for processing and forwarding user data. In traditional networks, it focuses on data packaging, forwarding, traffic management, and tunnel maintenance to ensure reliable end-to-end transmission of data packets. In contrast, the user plane in intelligent complex systems has been significantly expanded. It not only supports the transmission of service data but also carries environmental awareness information. In addition, the user plane can carry inputs, intermediate results, or inference results related to AI tasks \cite{yang2023task}. Furthermore, to accommodate diverse services, the user plane can flexibly adopt multiple data processing and forwarding strategies, including service awareness, path selection, computation-oriented data offloading and forwarding, and traffic control. 


\subsubsection{Data Plane}
The data plane is used to separate data resources from application service logic and serves as the foundational functional plane in intelligent complex systems. In traditional networks, data typically depends on the data forwarding process and lacks independent data management and service mechanisms. However, in intelligent complex systems, data has become a key resource supporting intelligent decision-making, model updates, and task processing. Consequently, intelligent networks require introducing a dedicated data plane to uniformly collect, clean, store, share, and orchestrate multi-source heterogeneous data, while providing data services to the control plane, user plane, computation plane and intelligence plane. Furthermore, as AI workflows in intelligent complex systems need to process massive data, the data plane manages data objects such as training samples, model inputs and outputs, intermediate results, model parameters, and gradients, thereby providing fundamental support for intelligent complex systems \cite{yuan20256g}.

\subsubsection{Computation Plane}

The computation plane is responsible for providing computational power to intelligent complex systems, serving as a critical foundation for task execution, model inference, and intelligent collaboration. It enables unified scheduling and management of computing resources across all planes to support functions such as feature extraction, distributed collaborative computing, and other computation-intensive tasks.

\subsubsection{Intelligence Plane}

The intelligence plane is primarily responsible for learning, inference, prediction, decision-making, and self-optimization, serving as the core enabler of comprehensive intelligence across complex systems \cite{cheng2025integrated}. Through interaction with other planes, it performs functions such as model deployment, training, inference, feedback, and evaluation, ultimately enhancing environmental adaptability and task execution capabilities within intelligent complex systems.

\subsubsection{Security Plane}
The security plane provides unified and trusted safeguards for intelligent complex systems, serving as a key component of cybersecurity, data security, model security, and cross-plane collaborative security \cite{meng2026intellicise}. In traditional networks, security objectives focus on authenticating identities and protecting communication links and user data from unauthorized access or corruption \cite{cheng2026apeg,meng2025survey2,liu2026optimization}. In contrast, within intelligent complex systems, the scope of security extends further to encompass data, models, knowledge, tasks, and more \cite{meng2025secure,nguyen2021security}. Moreover, as multi-plane architectures increasingly demand frequent interaction and collaboration between planes, the cross-plane collaboration process itself also requires security protection. 


\subsection{Novel Information Flow Structures}

\begin{table}
	\centering
	\caption{Comparison of Information Flows in Traditional and Intellicise Networks}
	\label{tab:information_flows}
	\renewcommand{\arraystretch}{1.35}
	\setlength{\tabcolsep}{5pt}
	
	\providecommand{\gc}{\cellcolor{gray!15}}
	
	\begin{adjustbox}{max width=\linewidth}
\begin{tabular}{
		|>{\raggedright\arraybackslash}m{2cm}
		|>{\raggedright\arraybackslash}m{5cm}
		|>{\raggedright\arraybackslash}m{10cm}|
	}
		\hline
		\textbf{Information Flow}
		& \textbf{Definition}
		& \textbf{Comparison with Traditional Networks}
		\\ \hline
		
		\gc Data Flow
		&
		\gc Carries raw observations and system states.
		&
		\gc $\bullet$ \textbf{Traditional:} Carries packets, symbols, or bit streams.\par
		$\bullet$ \textbf{Intellicise:} Extends to heterogeneous data generated by perception, communication, computation, and control across different layers.
		\\ \hline
		
		Knowledge Flow
		&
		Carries high-level semantic and relational information extracted and organized from transmitted data.
		&
		$\bullet$ \textbf{Traditional:} Lacks an explicit mechanism for knowledge transmission and sharing.\par
		$\bullet$ \textbf{Intellicise:} Transforms data into structured and relational knowledge, enabling consistent understanding across nodes and domains.
		\\ \hline
		
		\gc Model Flow
		&
		\gc Carries intelligence capabilities, including model architectures, parameter updates, and lightweight adaptation modules.
		&
		\gc $\bullet$ \textbf{Traditional:} Lacks an explicit mechanism for distributing intelligence capabilities.\par
		$\bullet$ \textbf{Intellicise:} Distributes and updates models across nodes to support collaborative intelligence and continuous adaptation.
		\\ \hline
		
		Task Flow
		&
		Carries task-oriented execution information and feedback across application, scheduling, and evaluation processes.
		&
		$\bullet$ \textbf{Traditional:} Focuses on data forwarding without an explicit task-level abstraction.\par
		$\bullet$ \textbf{Intellicise:} Translates application intents and resource requirements into executable strategies and refines them through task feedback.
		\\ \hline
		
	\end{tabular}
\end{adjustbox}
\end{table}


Within the proposed intellicise-network-based architecture, the paradigm of information exchange fundamentally evolves from conventional single-dimensional, bit-level transmission pipelines into a multi-dimensional mechanism grounded in a cross-domain fused structure \cite{zhang2024intellicise}. The proposed architecture systematically defines the information flow structures as data flow for data transmission and state updating, knowledge flow for knowledge expression and cognitive reasoning, model flow for intelligent capability supplementation, and task flow for task-oriented action and feedback. These four information flows interact to enable intelligent complex systems to achieve real-time perception, dynamic resource orchestration, and lifelong cognitive evolution, which are summarized in Table \ref{tab:information_flows}.

\subsubsection{Data Flow}

Data flow is the most fundamental information flow in intelligent complex systems. In traditional communication networks, it primarily refers to the transmission of packets, symbols, or bit streams \cite{yang2022semantic, gunduz2022beyond}. 
In intelligent complex systems, however, data flow is further broadened to encompass heterogeneous data generated by perception, communication, computation, and control across the entire cross-layer organizational framework. It includes externally acquired perceptual information at the perception and cognition layer, channel states and service traffic at the access and transmission layer, computing states and system interaction information at the interconnection and computing layer, and task execution states and control feedback at the decision and control layer. These data serve as the primary input for subsequent semantic extraction, knowledge construction, model update, and task execution.

\subsubsection{Knowledge Flow}
Knowledge flow is generated by extracting and organizing the inherent meaning and logical relations embedded in transmitted information.
In contrast to data flow, which mainly carries raw observations and system states, knowledge flow focuses on high-level semantic and relational information, including environmental knowledge, content knowledge, and control knowledge. Such knowledge can typically be formalized using dynamic knowledge graphs and retrieval-augmented generation \cite{fan2025kgrag}.
This shared knowledge provides a common background that helps different entities interpret information and understand system states in a consistent manner \cite{zhang2026towards}. 
Within intelligent complex systems, knowledge flow further connects knowledge distributed across different nodes, devices, and domains, enabling heterogeneous information to be organized under a shared semantic structure, thereby supporting cross-domain understanding and reasoning. 

\subsubsection{Model Flow}
Within intelligent complex systems, model flow not only provides the capabilities required by different layers but also delivers updated model parameters and structures to adapt to tasks and environments. Specifically, it dynamically distributes model architectures, such as Large Language Models (LLMs) and Large Vision Models (LVMs) \cite{jiang2026large}, along with localized updates like neural network weights, gradients, or lightweight parameter-efficient fine-tuning modules \cite{meng2026secure}.
The introduction of model flow further allows intelligent complex systems to deliver intelligence capabilities rather than merely data or knowledge, enabling them to improve generalization and support collaborative optimization. Through mechanisms such as collaborative edge inference and federated learning \cite{meng2025survey}, model flow enables locally updated model parameters or lightweight adaptation modules to be shared among distributed nodes, thereby linking intelligence capabilities with task requirements in intelligent complex system.

\subsubsection{Task Flow}
Task flow refers to the circulation of task-oriented execution and feedback in intelligent complex systems, focusing on how applications are executed across different domains and scenarios. Specifically, it carries task priority and resource consumption information to support the formulation of executable strategies for multi-scenario applications \cite{zhang2025resource}. Based on feedback collected during execution, task flow further refines these strategies and coordinates resource allocation among different tasks, thereby ensuring successful completion of complex scenarios. Thus, task flow provides task-level guidance for intelligent complex systems. It links application execution with feedback control, enabling the system to adjust resource allocation and service strategies according to task importance, which supports a reliable execution loop across diverse scenarios.

In general, the cross-layer organization achieves information exchange among the four layers, which is implemented through the vertical circulation of data, knowledge, model, and task flows, together with the horizontal coordination provided by the six functional planes. Data and knowledge generated at the perception and cognition layer can be used by the transmission, computing, and decision functions according to task requirements. Conversely,  task objectives and control policies generated at the decision and control layer guide semantic extraction, transmission configuration, and computing resource allocation. Guided by the cross-fused theories,information flows connect these operations across layers, while the functional planes coordinate the capabilities required to carry them out. This cross-layer organization reduces repeated information processing, supports joint optimization of communication, computing, and control resources, and enables the system to adapt to changing tasks and environmental conditions.

\section{Key Enabling Technologies for Intellicise Network-enhanced Intelligent Complex Systems}
\label{section4}

\begin{table}
	\centering
	\caption{Summary of Key Enabling Technologies and Their Evolutionary Objectives}
	\label{tab:key_technologies_overview}
	\renewcommand{\arraystretch}{1.5}
	\setlength{\tabcolsep}{5pt}
	
	\providecommand{\gc}{\cellcolor{gray!15}}
	
	\begin{adjustbox}{max width=\linewidth}
\begin{tabular}{
		|>{\raggedright\arraybackslash}m{4.5cm}
		|>{\raggedright\arraybackslash}m{8.3cm}
		|>{\raggedright\arraybackslash}m{4.2cm}|
	}
		\hline
		\textbf{Key Technology}
		& \textbf{Description}
		& \textbf{Objective}
		\\ \hline
		
		\gc From Semantic Extraction to Intent Understanding
		&
		\gc \textbf{Semantic Extraction:} Extracts task-relevant semantic information from multi-modal inputs.\par 
        \textbf{Intent Understanding:} Infers the underlying intents of users, agents, or systems.
		&
		\gc Shifts intelligent complex systems from passive response toward active understanding.
		\\ \hline
		
		From Heterogeneous Resource Integration to Self-configuration and Self-optimization
		&
		\textbf{Heterogeneous Resource Integration:} Integrates cross-domain and distributed heterogeneous resources.\par
        \textbf{Self-configuration and Self-optimization:} Leverages unified pooling, flexible scheduling, and intelligent orchestration for self-evolving.
		&
		Enables autonomous configuration, continuous optimization, and self-evolving network operation.
		\\ \hline
		
		\gc From Generative AI to Agentic AI
		&
		\gc \textbf{Generative AI:} Generates new content and system states from prompts or contextual conditions through deep generative models.\par
		\textbf{Agentic AI:} Extends generative capabilities with autonomous reasoning, planning, decision-making, tool use, and action execution.
		&
		\gc Evolves AI from a content generator and conversational assistant into an autonomous actor.
		\\ \hline
		
		From Embodied AI to Symbodied AI
		&
		\textbf{Embodied AI:} Perceives and interacts with the physical environment through embodied entities, learning and executing tasks based on environmental feedback.\par
		\textbf{Symbodied AI:} Integrates human cognition, intention, and physical capabilities into joint human--AI decision-making and execution.
		&
		Enables cognitive and physical symbiosis between humans and intelligent systems.
		\\ \hline
		
	\end{tabular}
\end{adjustbox}
\end{table}

The proposed architecture requires key enabling technologies to transform its cross-layer organization, multi-functional planes, and multidimensional information flows into practical system capabilities. In this section, the technological evolution from semantic extraction to intent understanding, from heterogeneous resource integration to self-configuration and self-optimization, from generative AI to agentic AI, and from embodied AI to symbodied AI is discussed as follow. The description and objective of key technologies are summarized in Table \ref{tab:key_technologies_overview}.

\subsection{From Semantic Extraction to Intent Understanding}

Semantic extraction provides the fundamental for efficient information representation and transmission in intelligent complex systems, while intent understanding further enables information processing to adapt to dynamic service and task requirements.By interpreting current objectives and receiver requirements, intent understanding guides which extracted semantics are prioritized and how they are represented and delivered. This evolution allows the system to move from what information is meaningful to determining what information is important \cite{lu2026important}. Representative technologies and their contributions are summarized in Table~\ref{tab:system_summary1}.

\begin{table}
	\centering
	\caption{Summary of Key Enabling Technologies: From Semantic Extraction to Intent Understanding in Intellicise Networks}
	\label{tab:system_summary1}
	\renewcommand{\arraystretch}{1.4}
	
	\providecommand{\gc}{\cellcolor{gray!15}}
	
	\begin{adjustbox}{max width=\linewidth}
\begin{tabular}{|m{2.7cm}|m{4cm}|m{5cm}|m{4.5cm}|}
		\hline
		\textbf{Category} 
		& \textbf{Sub-category} 
		& \textbf{Description} 
		& \textbf{Contributions} 
		\\ \hline
		
		Semantic Extraction 
		& \gc JSCC-based Semantic Extraction \cite{bourtsoulatze2019deep,dai2022nonlinear,yang2024swinjscc}
		& \gc Uses deep neural networks to extract latent semantic features from source data and encode them into semantic representation.
		& \gc Builds the representation basis for semantic transmission, supporting robust semantic delivery.
		\\ \cline{2-4}
		
		& Generative Prior-based Semantic Extraction \cite{lokumarambage2023wireless,grassucci2026generative,ren2025generative}
		& Employs generative models as priors to construct semantic conditions that guide synthesizing semantically consistent content.
		& Supports low-payload semantic transmission by leveraging generative priors for semantically consistent recovery.
		\\ \hline
		
		Intent Understanding 
		& \gc Intent Recognition and Representation \cite{zhang2026towards,sohrabi2016plan,thomas2022neuro}
		& \gc Recognizes intents from requests, behaviors, and contexts, and represents them in system-understandable forms.
		& \gc Provides the intent basis for clarifying user requirements, receiver demands, task purposes, and expected effects.
		\\ \cline{2-4}
		
		& Intent-Aware Critical Information Processing \cite{ye2025user,huang2025generalized,men2026video,tang2025semantic}
		& Identifies intent-critical information and supports prioritized processing and delivery.
		& Reduces redundant processing and transmission while improving the efficiency of information interaction.
		\\ \cline{2-4}
		
		& \gc Intent-Guided Strategy Adaptation \cite{liu2025receiver,jiang2026intention,chen2025goal}
		& \gc Adapts processing, transmission, scheduling, resource usage, and service-response strategies based on interpreted intents.
		& \gc Aligns system strategies with dynamic intents, improving service effectiveness and resource utilization.
		\\ \hline
		
	\end{tabular}
\end{adjustbox}
\end{table}

\subsubsection{Semantic Extraction for Intelligent Complex Systems}

Semantic extraction extracts semantic information from source data and constructs compact semantic representations to enable semantic communication \cite{xie2021deep}. In intelligent complex systems, heterogeneous sensors, terminals, edge nodes, and control modules continuously generate multi-modal observations, environmental states, service data, and task feedback. The continuous growth of this data expands the information space and increases the complexity of transmission and processing. By abstracting source data into semantic representations, semantic extraction reduces reliance on raw-data exchange and allows intelligent complex systems to organize information interaction at the semantic level under limited communication resources. Specifically, semantic extraction can be categorized into two technical approaches:

\begin{figure}
\centering
\includegraphics[width=0.8\linewidth,height=.58\textheight,keepaspectratio]{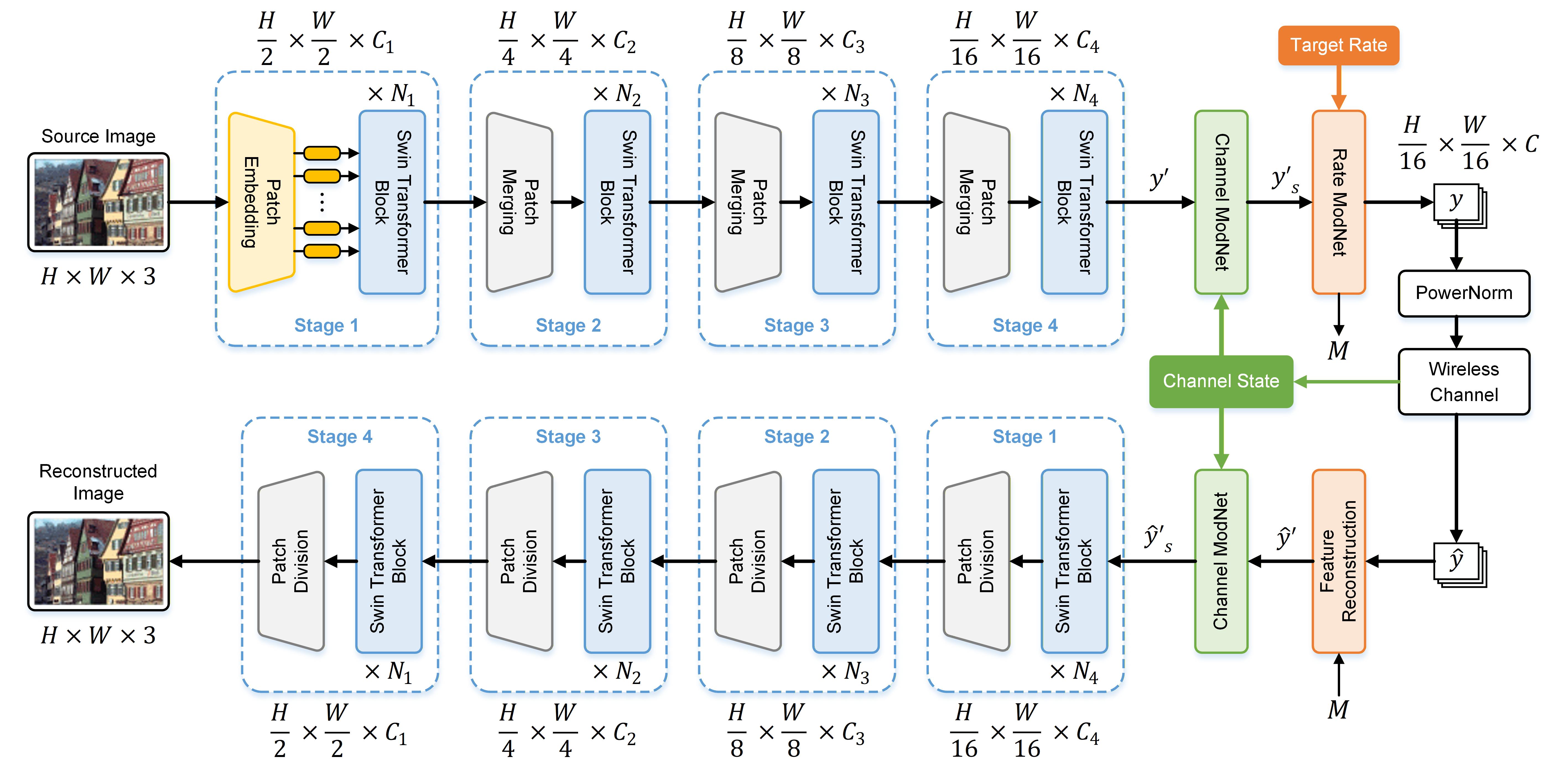}
\caption{Illustration of the SwinJSCC framework, where Swin Transformer-based JSCC extracts and transmits semantic representations for image semantic communication \cite{yang2024swinjscc}.}
\label{fig_semantic}
\end{figure}

\begin{itemize}
\item \textbf{JSCC-based Semantic Extraction:} 
Joint source-channel coding (JSCC)-based semantic extraction integrates semantic feature extraction, source coding, and channel transmission into a unified end-to-end learning framework \cite{bourtsoulatze2019deep,dai2022nonlinear}. Deep neural networks jointly learn source semantic representations and channel mappings, enabling source data to be directly mapped into robust semantic representation that preserve semantic information under channel distortion \cite{yang2023witt}. As shown in Fig. \ref{fig_semantic}, the core of this approach lies in jointly optimizing semantic extraction and physical transmission rather than treating them as independent procedures \cite{yang2024swinjscc}. For intelligent complex systems, this approach supports robust semantic-level delivery of heterogeneous source data under limited and time-varying wireless resources.

\item \textbf{Generative Prior-based Semantic Extraction:}
Generative prior-based semantic extraction employs generative models as shared semantic priors for semantic communication. In this approach, source data are transformed into compact semantic conditions, which then guide the generative model to synthesize semantically consistent content \cite{lokumarambage2023wireless,grassucci2026generative}. By leveraging the generation capability of shared semantic priors, semantic communication reduces transmission payload while maintaining semantic consistency with compact semantic conditions \cite{ren2025generative}. For intelligent complex systems with massive heterogeneous data, this approach provides an efficient means of supporting semantic interaction.
\end{itemize}

Semantic extraction provides a fundamental capability for intelligent complex systems to transition from raw-data exchange to semantic information interaction. It reduces unnecessary transmission while preserving the semantic information required for upper-layer intelligence and task-oriented operation \cite{xie2021deep,weng2023deep}. This capability is especially important in intelligent complex systems, where interactions among heterogeneous data and diverse service requirements expand the information space and increase processing complexity. By maintaining information interaction at the semantic level, semantic extraction enables the intelligent complex system to organize transmission and processing around task-relevant semantics, thereby reducing redundant information and supporting more efficient semantic communication, reasoning, and decision-making.

\subsubsection{Intent Understanding for Intelligent Complex Systems}

Semantic extraction constructs compact semantic representations as the basis for semantic-level transmission, processing, and reasoning in intelligent complex systems. However, for heterogeneous service requirements and dynamic task contexts, the system should further understand the intent behind information interaction. In intelligent complex systems, multiple users, services, and agents may coexist in the same networked environment. Their service and task requirements are usually heterogeneous and may change with service objectives, task states, environmental conditions, interaction contexts, and feedback.
Intent understanding addresses this need by interpreting dynamic requirements from requests, observed behaviors, and interaction contexts as system-understandable intents \cite{zhang2026towards}. These intents then guide critical information processing and strategy adaptation.

\begin{itemize}
\item \textbf{Intent Recognition and Representation:}
Intent understanding first interprets current service and task requirements and transforms them into system-understandable intent representations. When service requests, user queries, or task descriptions provide direct intent information, the corresponding intents can be identified from these inputs \cite{ye2025user,huang2025generalized}. In the absence of sufficient direct intent information, intents can also be inferred from observed behaviors, environmental states, interaction contexts, and feedback \cite{sohrabi2016plan,pereira2020landmark,dann2023multi,thomas2022neuro}. The intent representations can characterize aspects such as service objectives, task contexts, receiver requirements, and expected effects associated with information interaction. By organizing heterogeneous requirements into structured intent representations, they provide the basis for intent-critical information processing and strategy adaptation.

\item \textbf{Intent-Aware Critical Information Processing:}
Based on intent representation, information processing can be organized around intent-critical information. In intelligent complex systems, the intent relevance of information may change with service objectives, task states, and feedback. Treating all information with the same priority may therefore introduce unnecessary communication and computation costs. Intent understanding enables the system to identify information that is more closely related to current requirements and to support prioritized processing and delivery \cite{ye2025user,huang2025generalized,lu2025multi,liu2026communicate}. It also supports differentiated processing of intent-related information \cite{men2026video,tang2025semantic}. In this way, intent helps focus communication and computation resources on information that is more meaningful to current services and tasks.

\item \textbf{Intent-Guided Strategy Adaptation:}
Intent understanding also supports strategy adaptation for information interaction. Once the system understands current intents, it can adjust processing modules, representation forms, transmission strategies, scheduling policies, and resource usage according to changing requirements. Such adaptation allows information interaction to reflect service requirements and interaction contexts \cite{liu2025receiver,jiang2026intention,chen2025semantic}. It can also align resource usage with task utility and service requirements and support shared-goal alignment in multi-entity settings \cite{strinati2024goal,chen2025goal}. For intelligent complex systems, this supports adaptive information processing, resource usage, and service response under heterogeneous and changing requirements.
\end{itemize}



\subsection{From Heterogeneous Resource Integration to Self-configuration and Self-optimization}
Heterogeneous resource integration provides the resource foundation for communication, sensing and network operation in intelligent complex systems. Building on this foundation, self-configuration uses the states of integrated resources to autonomously establish suitable network settings, while self-optimization adjusts these settings in response to task feedback and environmental changes.
Representative technologies and their contributions are summarized in Table~\ref{tab:system_summary2}.

\begin{table}
	\centering
	\caption{Summary of Key Enabling Technologies: From Heterogeneous Resource Integration to Self-configuration and Self-optimization in Intellicise Networks}
	\label{tab:system_summary2}
	\renewcommand{\arraystretch}{1.4}
	
	\providecommand{\gc}{\cellcolor{gray!15}}
	
	\begin{adjustbox}{max width=\linewidth}
\begin{tabular}{|m{2.7cm}|m{4cm}|m{5cm}|m{4.5cm}|}
		\hline
		\textbf{Category} 
		& \textbf{Sub-category} 
		& \textbf{Description} 
		& \textbf{Contributions} 
		\\ \hline
		
		Heterogeneous Resource Integration 
		& \gc Mathematical Model Optimization \cite{zhang2025resource, hu2024exploiting}
		& \gc Formulates joint optimization frameworks by mapping heterogeneous resources to semantic metrics.
		& \gc Clarifies the fundamental paradigm for resource mapping and establishes theoretical performance bounds.
		\\ \cline{2-4}
		
		& AI-Driven Learning Optimization \cite{lv2025resource, xu2025heterogeneous, pellejero2025agentic, li2025federated}
		& Utilizes adaptive model training and agentic AI to learn allocation policies and synergize multi-modal features.
		& Enables real-time, model-free optimization in highly dynamic environments.
		\\ \hline
		
		Self-Configuration and Self-Optimization 
		& \gc Edge SLM Automation \cite{lwin2026performance, lira2026network}
		& \gc Deploys SLMs at the edge to parse operational requirements and synthesize valid network topologies.
		& \gc Reduces computational overhead and latency for dynamic, automated network setup.
		\\ \cline{2-4}
		
		& Cognitive Loop Evolution \cite{zhang2026towards}
		& Implements a continuous perception-memory-reasoning-action loop for network self-evolution.
		& Ensures long-term resilience and continuous adaptation to environmental shifts.
		\\ \cline{2-4}
		
		& \gc Decentralized Agentic Collaboration \cite{xiao2026sanet}
		& \gc Deploys multi-agent frameworks to dynamically optimize cross-layer resources.
		& \gc Converges on Pareto-optimal states efficiently without requiring global state observability.
		\\ \hline
		
	\end{tabular}
\end{adjustbox}
\end{table}

\subsubsection{Heterogeneous Resource Integration for Intelligent Complex Systems}
To establish the foundation of intelligent complex systems, the intellicise networks emphasize a holistic convergence of upcoming technological trends to meet the rigorous performance requirements of diverse autonomous services \cite{saad2019vision}. Unlike traditional networks that allocate only physical communication elements, intellicise network-based intelligent complex systems introduce a highly coupled heterogeneous resource pool, which includes:
\begin{itemize}
	\item \textbf{Communication Resources:} These refer to the wireless resources employed by base stations (BSs) or user devices for information transmission, including bandwidth, transmit power, and subchannel allocation strategies\cite{nie2025heterogeneous, 11683888}.
	\item \textbf{Computing Resources:} This category encompasses the computational frequency or capacity of CPUs and GPUs available on both the BS and user sides, which are essential for executing complex semantic extraction and reconstruction algorithms \cite{liu2023efficient,wang2026optimization}.
	\item \textbf{Network Resources:} These include the hardware and logical resources of the bearer network and core network that enable end-to-end data transmission and network control. They comprise bearer plane resources, control plane resources and network slicing resources \cite{duan20236g}.
	\item \textbf{Storage Resources:} These hardware resources at edge servers or BSs are used to cache computing tasks, background KBs, and so on to enhance real-time analytics \cite{xu2023cloud}.
	\item \textbf{Sensing Resources:} These refer to perception-related hardware and capabilities deployed on BSs, user terminals, and edge nodes, including sensors, sampling frequency, sensing bandwidth, and environmental data acquisition capacity. They capture multi-dimensional physical information such as location, motion status, and ambient signals, providing raw perceptual inputs for intelligent complex systems\cite{feng2022joint, Sun2026TWC}.
\end{itemize}

\begin{figure}
\centering
\includegraphics[width=0.8\linewidth,height=.58\textheight,keepaspectratio]{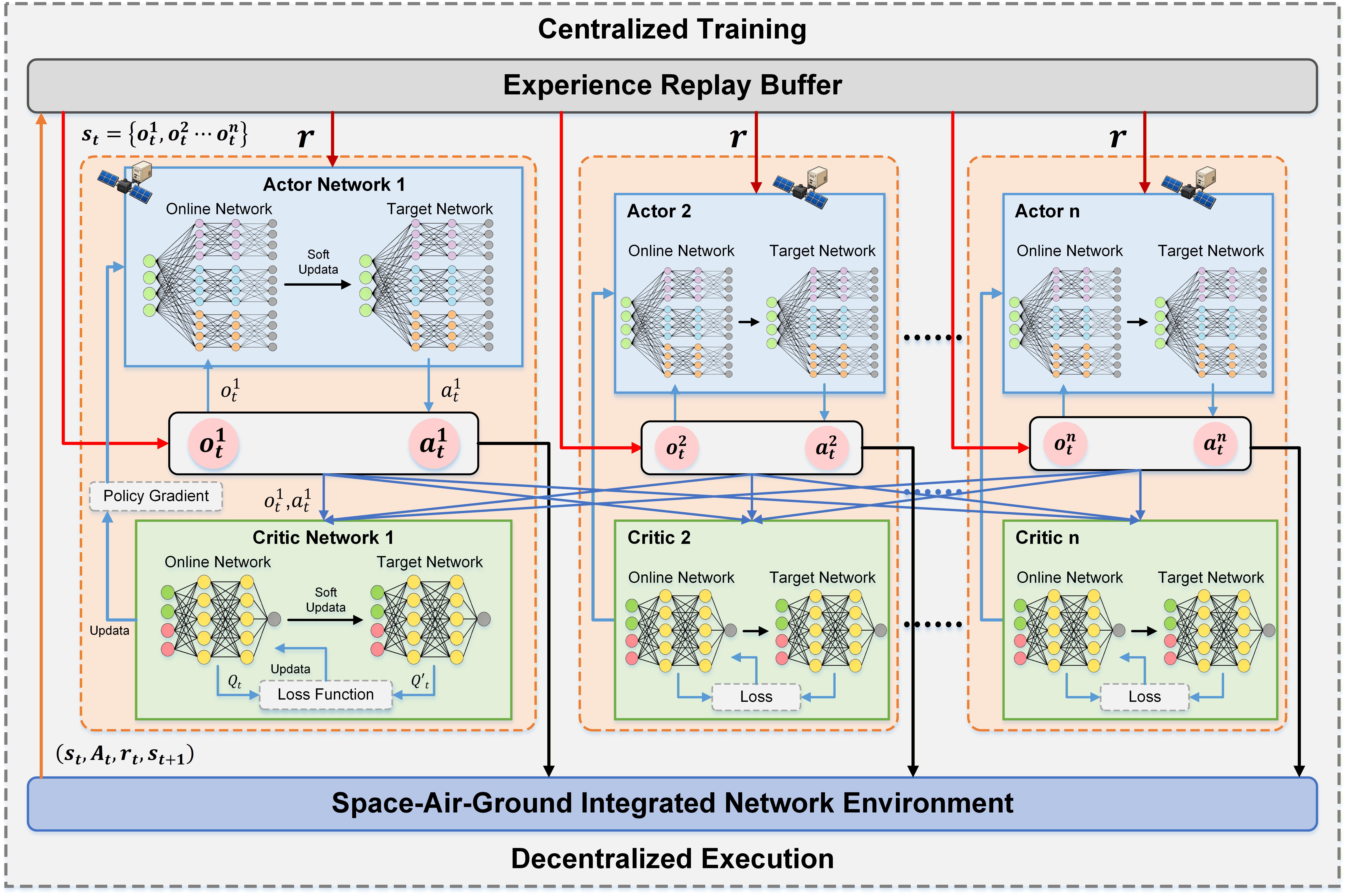}
\caption{Illustration of the Centralized Training with Decentralized Execution (CTDE)-MADDPG-based resource allocation framework, where LEO satellite agents learn joint offloading and heterogeneous resource allocation decisions through centralized training and decentralized execution \cite{xu2025heterogeneous}.}
\label{fig_resource}
\end{figure}

The integration of these heterogeneous resources is not a simple summation but a deep functional coupling aimed at maximizing the overall utility of intelligent complex systems. Given the massive device connectivity and multi-modal data streams in 6G, the integration paradigms have evolved systematically. Specifically, they can be categorized into the following two technical approaches:
\begin{itemize}
\item \textbf{Mathematical-Model Optimization:} Traditional approaches formulate resource allocation as a joint mathematical problem to derive exact, calculable theoretical performance bounds \cite{Sun2025CM}. In the context of intelligent complex systems, the core of this approach lies in constructing objective functions that map heterogeneous resource types to corresponding semantic performance metrics based on semantic similarity \cite{zhang2025resource}. This fundamental paradigm involves mapping resources to semantic performance, formulating objectives based on interdependencies, and deploying analytical optimization algorithms for efficient utilization. Furthermore, methods like hierarchical coded multi-task learning utilize complex network topology to establish optimal data distribution and resource mapping mathematically \cite{hu2024exploiting}.

\item \textbf{AI-Driven Learning Optimization:} In contrast to calculable mathematical models, highly dynamic and high-dimensional environments require continuous model training and adaptation \cite{xu2025transformer}. As shown in Fig. \ref{fig_resource}, this category unifies deep reinforcement learning (DRL) and Agentic architectures \cite{xu2025heterogeneous}. DRL methods, particularly proximal policy optimization (PPO) and dueling deep Q-networks (DDQN), enable agents to interact with environments to learn optimal allocation policies in a model-free manner. For instance, recent studies implement DDQN to dynamically orchestrate computing, communication, and sensing resources, demonstrating superior high-order bit error resilience \cite{lv2025resource, xu2025heterogeneous}. 
\end{itemize}
\subsubsection{Self-Configuration and Self-Optimization for Intelligent Complex Systems}

\begin{figure}
\centering
\includegraphics[width=0.7\linewidth,height=.58\textheight,keepaspectratio]{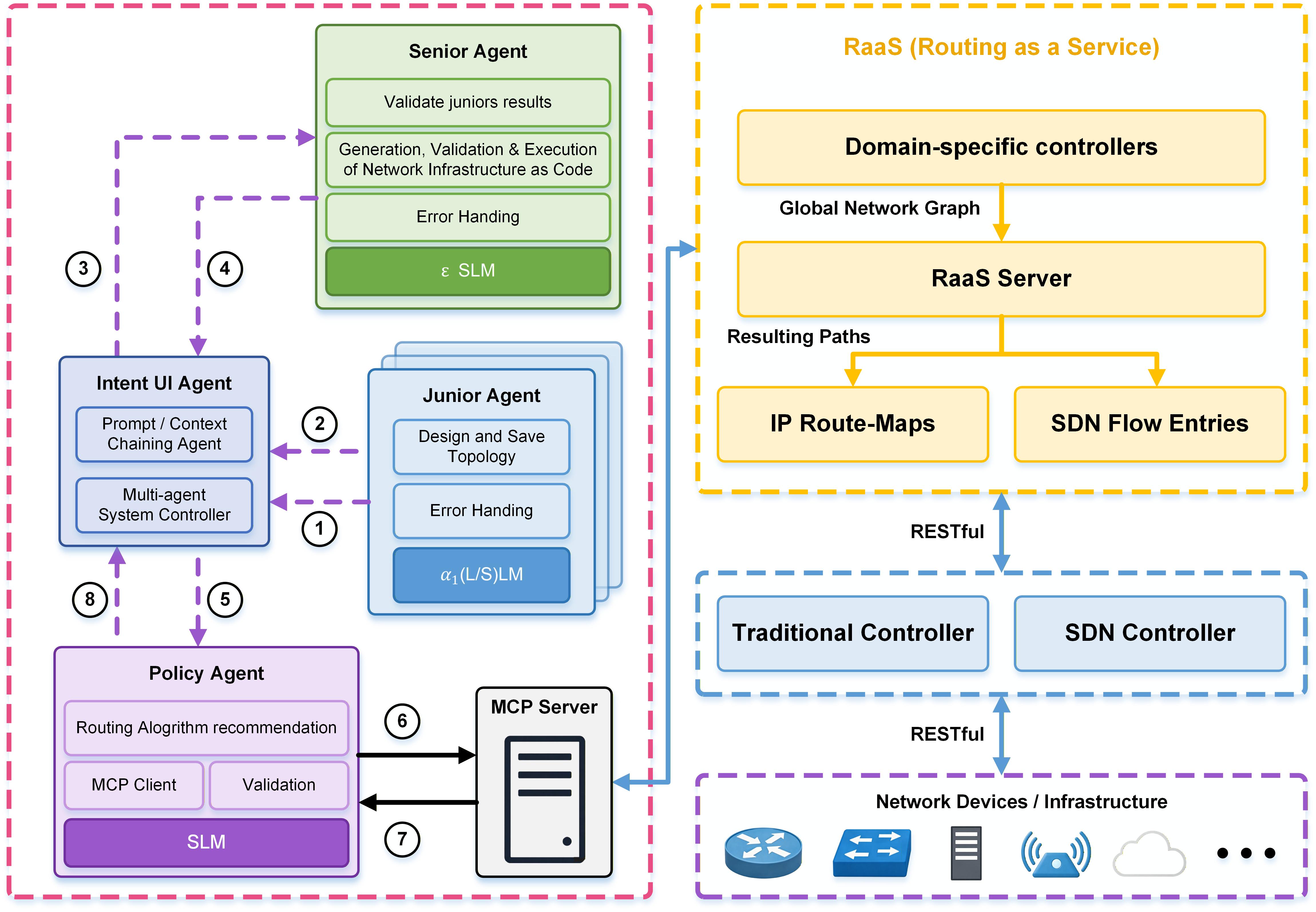}
\caption{Illustration of the intent-based network orchestration framework, where agents based on LLMs and SLMs automate topology configuration, validation, and RaaS-assisted routing optimization based on network intents for self-configuration and self-optimization \cite{lwin2026performance}.}
\label{fig_self}
\end{figure}

The progression from heterogeneous resource integration naturally extends into the unified lifecycle of self-configuration and self-optimization. In intelligent complex systems, the sheer scale and extreme heterogeneity of 6G deployments render early static, rule-based initialization methods obsolete \cite{peng2013self}. This lifecycle involves automated topology configuration and continuous optimization during operation.

\begin{itemize}
\item \textbf{Edge SLM Automation:}
For dynamic self-configuration, the network must autonomously establish its structural topology. To achieve this without incurring the high computational costs of cloud-based models, recent frameworks deploy parameter-efficient small language models (SLMs) directly at the edge, as shown in Fig. \ref{fig_self}. These SLMs autonomously parse high-level operational requirements to synthesize syntactically valid network topologies, achieving automated configuration with significantly lower latency \cite{lwin2026performance, lira2026network}.

\item \textbf{Cognitive Loop Evolution:}
Once initialized, the intelligent complex system must seamlessly transition to continuous self-optimization to maintain resilience against environmental fluctuations. This active adaptation shifts the network from initial orchestration to a self-evolving entity governed by a continuous perception-memory-reasoning-action cognitive loop \cite{zhang2026towards}. Through this loop, the system can continuously adjust its configurations and optimization strategies according to environmental changes and service requirements.

\item \textbf{Decentralized Agentic Collaboration:}
Within this transformative paradigm, decentralized frameworks deploy specialized AI agents that collaborate dynamically to optimize cross-layer resources. By interacting through the cognitive loop, these agents can converge upon Pareto-optimal resource states without requiring full global observability, continuously self-correcting to escape local optima \cite{xiao2026sanet}.
\end{itemize}

\subsection{From Generative AI to Agentic AI}
Generative AI enhances the representation, reconstruction, and prediction of complex system states, providing a powerful modeling foundation for intelligent complex systems. Moreover, agentic AI further extends these capabilities toward perception, sensing, memory and action by using these representations and predictions to plan actions, interact with network tools, and adjust decisions based on feedback, transforming AI from a content-generation capability into an operational intelligence capability. Representative technologies and their contributions are summarized in Table~\ref{tab:genai_agenticai_summary}.

\begin{table}[htbp]
	\centering
	\caption{Summary of Key Enabling Technologies: From Generative AI to Agentic AI in Intellicise Networks}
	\label{tab:genai_agenticai_summary}
	\renewcommand{\arraystretch}{1.4}
	
	\providecommand{\gc}{\cellcolor{gray!15}}
	
	\begin{adjustbox}{max width=\linewidth}
\begin{tabular}{|>{\centering\arraybackslash}m{1.8cm}|m{4.2cm}|m{5cm}|m{4.5cm}|}
		\hline
		\textbf{Category} 
		& \textbf{Sub-category} 
		& \textbf{Description} 
		& \textbf{Contributions} 
		\\ \hline
		
		\multirow{4}{=}{Generative AI}
		& \gc Multi-source State Learning \cite{zhou2026digtwinai}
		& \gc Learns system state distributions from multi-source observations for uncertainty-aware representation.
		& \gc Reduces semantic redundancy and enhances perception for intelligent complex systems.
		\\ \cline{2-4}
		
		& Digital Twin Synthesis \cite{zhou2026digtwinai}
		& Enhances digital twins by synthesizing long-tail events under limited sensing and bandwidth.
		& Improves model robustness and supports predictive operation for complex system management.
		\\ \cline{2-4}
		
		& \gc Semantic Encoder and Decoder \cite{gunduz2022beyond}
		& \gc Acts as semantic encoder and decoder, compressing task-relevant meaning into latent representations.
		& \gc Improves transmission efficiency and supports cross-layer semantic interaction.
		\\ \cline{2-4}
		
		& Network State Generation \cite{zhou2026digtwinai}
		& Generates network state samples including channel evolution and traffic patterns for robust planning and resource orchestration.
		& Provides data augmentation for learning-based control and improves resilience to distribution shifts.
		\\ \hline
		
		\multirow{4}{=}{Agentic AI}
		& \gc Intent Translation and Deployment \cite{zhang2026agenticsurvey, leivadeas2022survey}
		& \gc Translates user intents into executable task graphs and deploys operation through tool-based control interfaces.
		& \gc Enables verifiable intent deployment and controllable system operation.
		\\ \cline{2-4}
		

		& Closed-Loop Agentic Orchestration \cite{zhang2026agenticsurvey}
		& Monitors indicators and refines configurations to form perception-memory-reasoning-action loop.
		& Supports continuous self-configuration, self-optimization, and adaptive operation.
		\\ \cline{2-4}
		
		& \gc Specialized Agent Organization \cite{zhang2026agenticsurvey}
		& \gc Organizes specialized agents for perception, optimization, knowledge updating, and policy enforcement.
		& \gc Enhances scalability and resilience for distributed intelligent complex systems.
		\\ \cline{2-4}
		
		& Semantic Native Networking \cite{zhou2026digtwinai}
		& Relies on semantic native networking and shared knowledge structures to ensure reliable collaboration among agents.
		& Strengthens cross-domain coordination and supports stable closed-loop control in large-scale deployments.
		\\ \hline
		
	\end{tabular}
\end{adjustbox}
\end{table}


\subsubsection{Generative AI for Intelligent Complex Systems}
\begin{figure}
\centering
\includegraphics[width=\linewidth,height=.58\textheight,keepaspectratio]{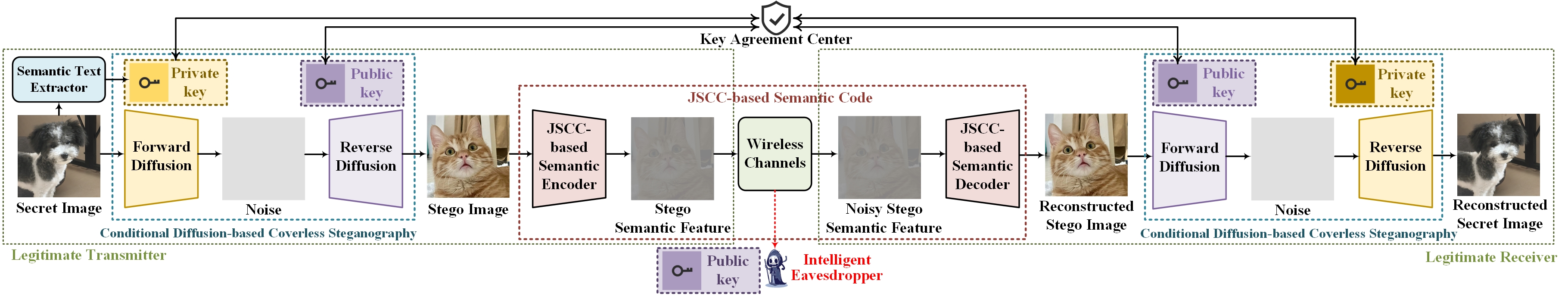}
\caption{Illustration of the SemSteDiff framework, where the generative diffusion model is employed to hide secret information into the generated image to achieve covert semantic communication \cite{gao2026semstediff}.}
\label{fig_generative}
\end{figure}

Generative AI provides a principled way to learn, represent, and synthesize high dimensional system states and their evolutions, which are difficult to capture using explicit models in open and dynamic environments \cite{zhou2026digtwinai,meng2026generative}. This capability helps intelligent complex systems handle incomplete observations and dynamic environments while supporting semantic interaction, thereby providing the modeling basis for higher-level autonomy.

\begin{itemize}
\item \textbf{Multi-source State Learning:}
For intelligent complex systems, massive multi-source data streams are collected by sensing devices, network nodes, and computing platforms, yet the observations are often sparse, delayed, and partially missing. By learning the underlying data distribution, generative models can reconstruct missing semantics, denoise corrupted observations, and provide calibrated uncertainty for downstream decision-making. This capability reduces the burden of transmitting redundant raw data, because the system can transmit compact semantic representations and recover detailed information only when required by a task \cite{wang2023improved, onsu2025semantic}.



\item \textbf{Digital Twin Synthesis:}
A central use of generative AI in intelligent complex systems is to enhance the digital twin pipeline. Digital twins require continuous synchronization between physical processes and their digital representations, and this synchronization is challenged by bandwidth limits, sensing noise, and long-tail events. Generative models can synthesize realistic system trajectories, generate rare failure cases, and augment training data for prediction and diagnosis. When combined with simulation and domain constraints, generative models help digital twins move beyond static mirroring and toward predictive and prescriptive operation \cite{zhou2026digtwinai}. 
Generative models can also act as compact world models that support fast what if analysis and scenario exploration, which is valuable for resource orchestration under uncertainty.

\item \textbf{Semantic Encoder and Decoder:}
Generative AI further enables semantic-level transformation rather than bit-level replication \cite{bastek2025physics, nguyen2025physix,gao2026semstediff}. 
Generative models can serve as semantic encoders and decoders that compress information into latent representations while preserving task utility, and they can generate plausible reconstructions conditioned on shared knowledge \cite{ren2025generative2}. This supports cross-layer semantic interaction, because semantic representations can be produced at the source, transported efficiently, and refined at the destination using KBs. Moreover, generative models can also support secure transmission. As illustrated in Fig.~\ref{fig_generative}, the SemSteDiff framework hides secret information into diffusion-generated images, allowing the generated content to preserve its normal semantic appearance while simultaneously serving as a covert carrier \cite{gao2026semstediff}. 

\item \textbf{Network State Generation:}
In addition, generative models can assist network modeling by generating channel states, traffic patterns, and topology evolution samples for robust planning \cite{zhou2026digtwinai}. These generated samples can complement sparse or incomplete observations and provide additional state scenarios for learning-based control. By modeling possible state evolutions under changing network conditions, network state generation improves robustness to distribution shifts and supports resource orchestration and decision-making under uncertainty.
\end{itemize}



\subsubsection{Agentic AI for Intelligent Complex Systems}

Agentic AI elevates generative AI into a closed-loop decision and control capability by integrating planning, tool use, and feedback-driven adaptation \cite{zhang2026agenticsurvey,  yan2025mas}. For intelligent complex sysytem, this link high-level reasoning with operational control, enabling the decisions and actions to be planned, executed, monitored, and adjusted under dynamic conditions. 

\begin{itemize}
\item \textbf{Intent Translation and Deployment:}
In intelligent complex systems, the core challenge is not only to understand semantics but also to transform intent into coordinated actions across heterogeneous resources. Agentic AI provides this transformation by decomposing high-level intents into executable task graphs, selecting actions under resource constraints, and interacting with external tools such as controllers, schedulers, KBs, and network management functions. This directly strengthens the decision and control layer, and it operationalizes the task flow by turning task objectives into measurable execution steps and feedback signals. In intent-based networking, natural language goals are translated into configuration policies and operational procedures \cite{wang2025intent}. Agentic AI can interpret intents, generate candidate configurations, validate them against constraints, and deploy them through control interfaces. 

\item \textbf{Closed-Loop Agentic Orchestration:}
After deployment, agentic AI can then monitor performance indicators and iteratively refine the configuration in response to traffic variation, mobility, and service changes. This forms a continuous perception-memory-reasoning-action loop that is consistent with the self-configuration and self-optimization targets of intellicise networks. Unlike static rule-based automation, agentic AI can generalize across scenarios by using knowledge retrieval and model adaptation, and it can coordinate across planes by jointly considering communication, computing, network, storage, and sensing resources. 
Beyond network configuration, the same closed-loop orchestration capability can also support security-sensitive communication services. As illustrated in Fig. \ref{fig_agentic}, agentic AI coordinates semantic understanding, adaptive coverless stego-content generation, and secret information recovery, thereby integrating covert transmission into the semantic communication process.

\begin{figure}
\centering
\includegraphics[width=\linewidth,height=.58\textheight,keepaspectratio]{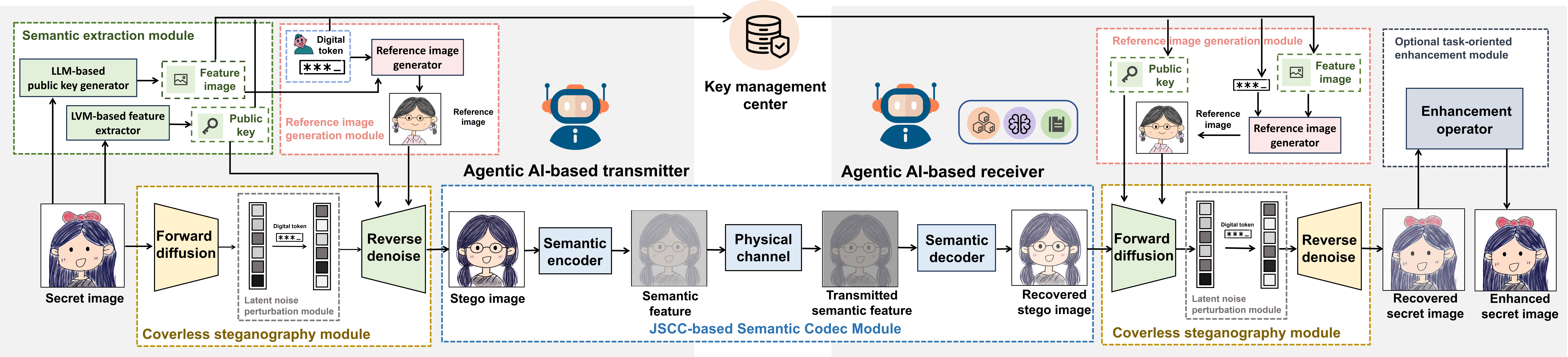}
\caption{Illustration of the agentic AI-enabled coverless semantic steganography communication framework, where agentic AI coordinates semantic understanding, adaptive stego-content generation, and secret information recovery to achieve secure transmission \cite{meng2026secure}.}
\label{fig_agentic}
\end{figure}

\item \textbf{Specialized Agent Organization:}
Because intelligent complex systems are inherently distributed, agentic AI must also support multi-agent collaboration. A single agent rarely has complete information or sufficient authority to manage cross-domain tasks. Multi-agent systems allow specialized agents to handle perception, resource negotiation, model adaptation, and security auditing in parallel, while maintaining a coherent global objective. Such specialization distributes decision-making responsibilities across different agents and supports scalable collaboration in large-scale intelligent complex systems.

\item \textbf{Semantic Native Networking:}
Specialized agents requires structured agent communication and coordination mechanisms that are compatible with semantic-native networking. The intellicise networks can provide an efficient substrate for such collaboration by transporting intent, state, and tool outputs as semantic information rather than raw messages, and by maintaining a shared knowledge structure that ensures consistent interpretation among agents. Through multi agent collaboration, the system gains resilience against partial failures and improves scalability \cite{chen2025goal}, which is essential for large scale intelligent complex systems such as industrial networks, transportation networks, and digital twin-enabled infrastructures.
\end{itemize}

\subsection{From Embodied AI to Symbodied AI}
Embodied AI grounds system intelligence in the physical environment through perception, decision-making, and action, enabling autonomous interaction under real-world constraints. Symbodied AI extends this paradigm toward deeper human-machine collaboration by incorporating human intent, cognition, and feedback into the intelligent decision loop, which shapes both the agent’s actions and the task-relevant information exchanged during execution. Representative technologies and their contributions are summarized in Table~\ref{tab:eai_sai_summary}.

\begin{table}
    \centering
    \caption{Summary of Key Enabling Technologies: From Embodied AI to Symbodied AI in Intellicise Networks}
    \label{tab:eai_sai_summary}
    \renewcommand{\arraystretch}{1.4}

    \providecommand{\gc}{\cellcolor{gray!15}}

    \begin{adjustbox}{max width=\linewidth}
\begin{tabular}{|m{2.2cm}|m{3.6cm}|m{5.4cm}|m{4.8cm}|}
        \hline
        \textbf{Category}
        & \textbf{Sub-category}
        & \textbf{Description}
        & \textbf{Contributions}
        \\ \hline

        \multirow{3}{*}{\shortstack[c]{Embodied AI}}
        & \gc Physical Interaction Loop \cite{duan2022survey, zitkovich2023rt, letaief2019roadmap}
        & \gc Enables agents to interact within physical perception-decision-execution loops under dynamic constraints.
        & \gc Provides the physical substrate and active probing for digital-physical network integration.
        \\ \cline{2-4}

        & Endogenous World Model \cite{ha2018world}
        & Internalizes physical laws within network logic to bypass pure statistical data.
        & Enables predictive interaction and filters redundant signals to suppress perceptual entropy.
        \\ \cline{2-4}

        & \gc Task-Critical Semantic Scheduling \cite{zhang2024intellicise, luo2022semantic, gunduz2022beyond, shao2021learning}
        & \gc Couples task-relevant semantic extraction with attention-guided resource allocation, shifting communication and scheduling goals from bit-accuracy to physical task success.
        & \gc Redefines network roles for mission-driven environment configuration and streamlines protocol hierarchies by merging semantic communication with task-aware scheduling.
        \\ \hline

        \multirow{2}{*}{\shortstack[c]{Symbodied AI}}
        & Human-Centric Symbiotic Alignment \cite{licklider2008man, driess2023palm, Lu2025SymbodiedAI, sun2025collabvla, ouyang2022training}
        & Embeds human first-person perception, intent, and feedback into the decision loop through collaborative VLA, self-reflective co-adaptation, and reinforcement learning from human feedback.
        & Transforms autonomous machines into symbiotic partners aligned with human cognitive mindsets and resolves intent safety issues for high-fidelity coordination.
        \\ \cline{2-4}

        & \gc Intent-Aware Cognitive Orchestration \cite{leivadeas2022survey, strinati20196g, strinati20216g, rexford2004network}
        & \gc Parses intent-driven, high-entropy data and introduces an endogenous cognitive mechanism above traditional control and transmission structures to guide resource allocation and collaborative reasoning.
        & \gc Drives feature-space resource allocation and extreme communication pruning based on attention saliency, and supports system-level entropy reduction through human-machine collaborative reasoning.
        \\ \hline

    \end{tabular}
\end{adjustbox}
\end{table}

\subsubsection{Embodied AI for Intelligent Complex Systems}

Embodied AI can be characterized by three intertwined capabilities that reshape how intelligent complex systems interact with the physical world: physical grounding through closed-loop interaction, predictive modeling via endogenous world models, and task-critical semantic scheduling.

\begin{figure}
\centering
\includegraphics[width=0.75\linewidth,height=.58\textheight,keepaspectratio]{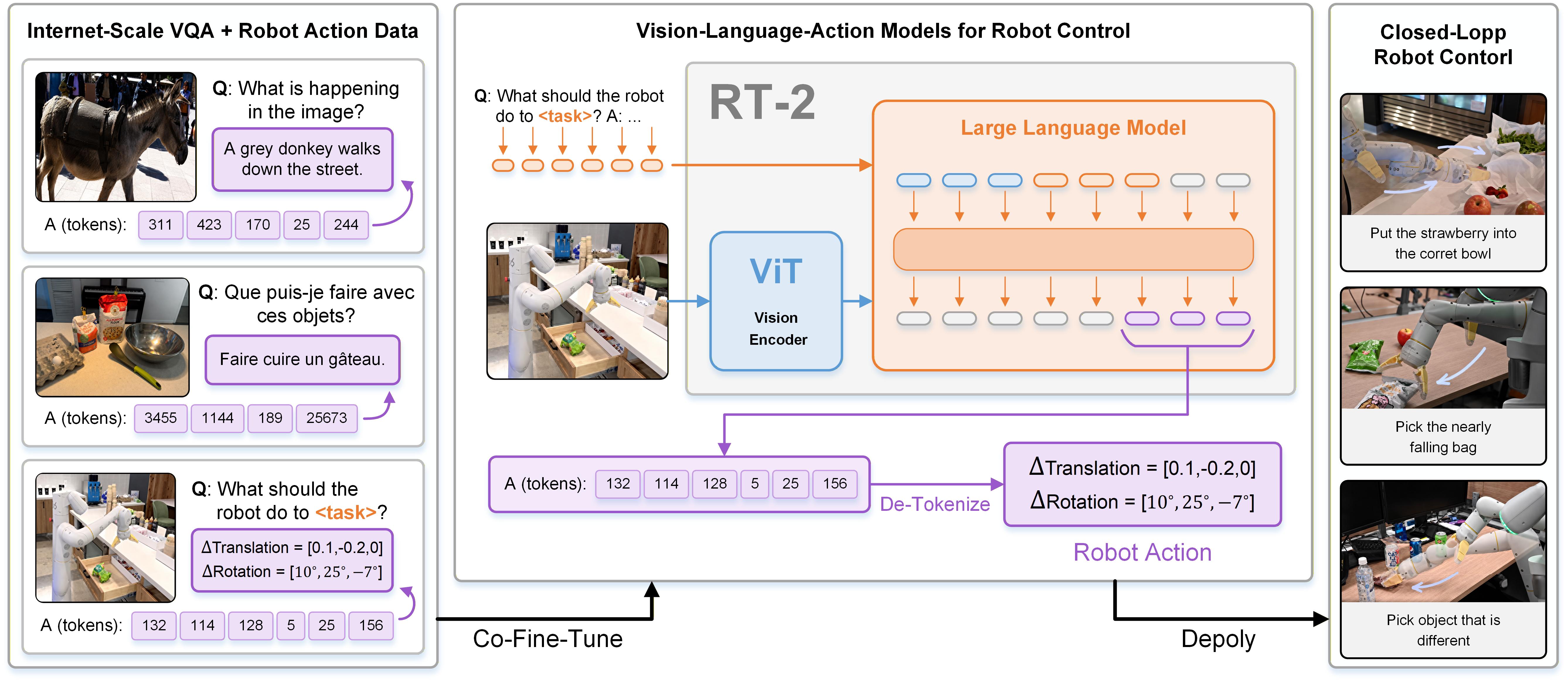}
\caption{Illustration of the RT-2 framework, where the model is co-fine-tuned on VQA and robot action data to map visual observations and language instructions into action tokens for closed-loop embodied control \cite{zitkovich2023rt}.}
\label{fig_embodied}
\end{figure}

\begin{itemize}
\item \textbf{Physical Interaction Loop:}
Embodied AI marks a shift from disembodied intelligence to a paradigm grounded in physical entities \cite{duan2022survey}. Rather than acting as passive information nodes, embodied AI agents operate within a closed perception-decision-execution loop \cite{zitkovich2023rt}, as shown in Fig. \ref{fig_embodied}. This allows them to interact directly with physical constraints such as dynamic characteristics, spatial poses, and positions. In intelligent complex systems, embodied AI serves a dual role \cite{letaief2019roadmap}: it extends communication to the pragmatic level and acts as an active probe that explores and verifies environmental states. By replacing passive data collection with active physical interaction, embodied AI provides the physical substrate needed to manage entropy growth in intelligent complex systems.

\item \textbf{Endogenous World Model:}
Embodied AI internalizes physical laws within the network's logic. The system therefore moves beyond pure statistical pattern matching and acquires an endogenous world model \cite{ha2018world}. This internalization enhances network intelligence, enabling task-oriented environment configuration and predictive interaction \cite{zhang2024intellicise}. Because the world model lets the network anticipate physical-state changes, it can filter physically impossible or redundant signals at the perceptual source, thereby controlling entropy \cite{luo2022semantic}.

\item \textbf{Task-critical Semantic Scheduling:}
Embodied AI couples semantic communication and resource scheduling rather than treating them as separate functions. The communication objective shifts from bit-level accuracy under the Shannon paradigm to maximizing the success rate of embodied control tasks \cite{gunduz2022beyond}. Attention mechanisms then allocate communication and computing resources to the semantic features that matter for the current physical task, such as spatial logic relationships and anomaly state signals, while suppressing irrelevant background noise \cite{shao2021learning}. This coupling streamlines protocol hierarchies and merges resource scheduling with physical task logic, enabling mission-driven environment configuration \cite{zhang2024intellicise, luo2022semantic}.
\end{itemize}

Together, the physical interaction loop, the endogenous world model, and task-critical semantic scheduling form a coherent embodied intelligence capability. They ground network intelligence in physical reality, anticipate environmental dynamics, and align communication and scheduling with task objectives. Embodied AI thereby transforms intelligent complex systems from passive data networks into active, task-aware physical agents that probe, model, and act within their environments.


\subsubsection{Symbodied AI for Intelligent Complex Systems}

Symbodied AI extends embodied execution toward human-machine cognitive alignment through two complementary capabilities: human-centric symbiotic alignment that couples perception, intent, and feedback, and intent-aware cognitive orchestration that turns intent into resource allocation and system-level reasoning.

\begin{itemize}
\item \textbf{Human-centric Symbiotic Alignment:}
Symbodied AI goes beyond physical execution to emphasize deep cognitive alignment and co-evolution between humans and machines. Instead of treating human input as an external command channel, it embeds first-person perception, intent, and feedback into the decision loop through collaborative vision-language-action (Co-VLA) infrastructures and self-reflective co-adaptation \cite{driess2023palm, sun2025collabvla, Lu2025SymbodiedAI}. Reinforcement learning from human feedback further aligns execution with human values and resolves intent-safety issues \cite{ouyang2022training}. 


\item \textbf{Intent-aware Cognitive Orchestration:}
As symbodied AI matures, it imposes a strong reverse pressure on the underlying intellicise networks. High-dimensional intent signals are parsed to guide both resource allocation and collaborative reasoning \cite{leivadeas2022survey, strinati20196g}. Resource management shifts from physical-link allocation to feature-space, attention-guided allocation, pruning communication according to the saliency of information relative to the human's current focus \cite{strinati20216g}. The endogenous cognition further supports human-machine collaborative reasoning and system-level entropy reduction \cite{rexford2004network}. Intent therefore becomes the organizing principle that couples cognition, communication, and control in symbiotic intelligent complex systems.
\end{itemize}

\begin{figure}
\centering
\includegraphics[width=0.75\linewidth,height=.58\textheight,keepaspectratio]{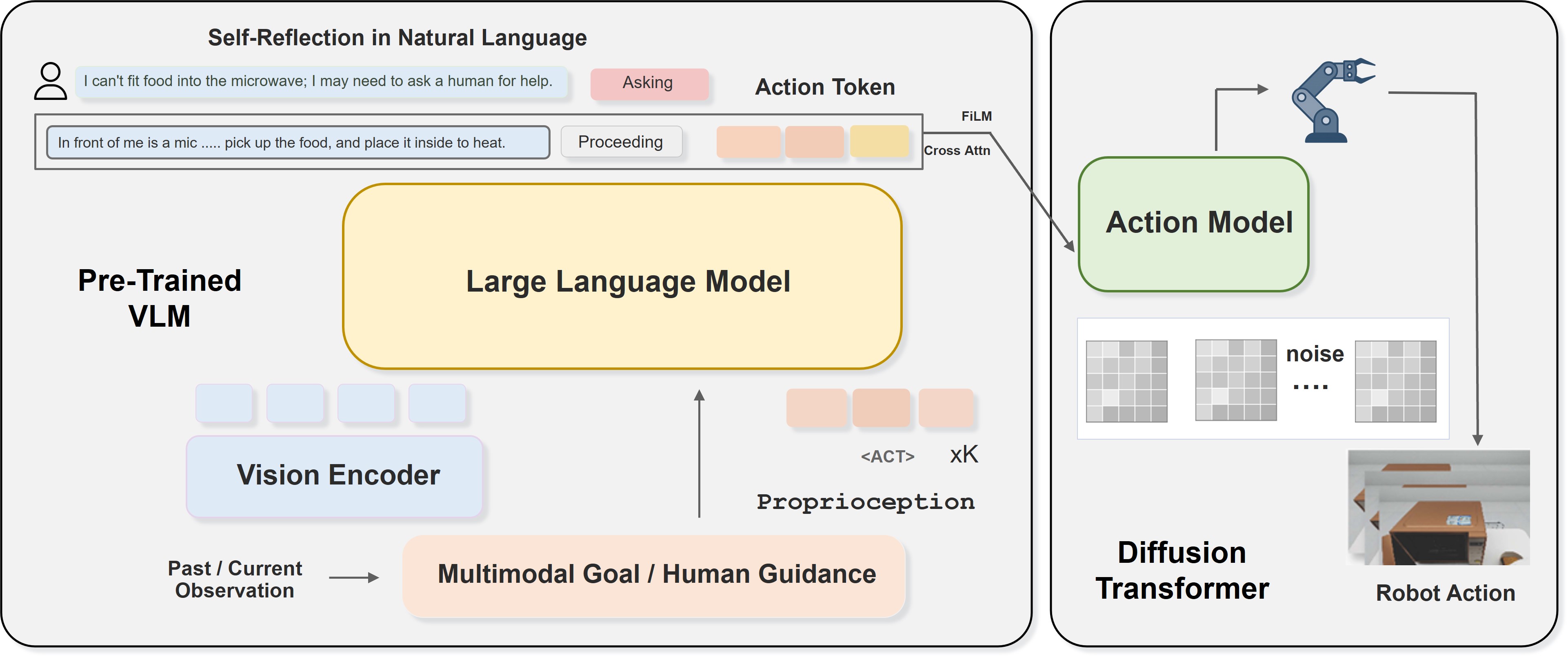}
\caption{Illustration of Symbodied AI through the CollabVLA framework, where self-reflective reasoning and human guidance jointly inform action generation, enabling the agent to adapt its behavior and collaborate with humans during physical task execution \cite{sun2025collabvla}.}
\label{fig_symbodied}
\end{figure}

Together, human-centric symbiotic alignment and intent-aware cognitive orchestration elevate embodied intelligence to human-machine co-evolution. They let intelligent complex systems interpret human intent, align actions with human values, and orchestrate network resources and reasoning around those intents. 


\subsection{Summaries and Lessons Learned}

This section reviews key enabling technologies that support intelligent complex systems in representing information, coordinating resources, modeling dynamic states, and interacting with physical and human environments. The following discussion summarizes lessons and open problems for building more reliable and adaptive intelligent complex systems.

\begin{itemize}
\item 
Semantic extraction enables semantic-level information interaction by constructing compact representations from heterogeneous source data and reducing redundant raw-data exchange \cite{dai2022nonlinear,yang2024swinjscc,ren2025generative}. Intent understanding interprets dynamic service and task requirements into intents, and guides information processing, transmission, scheduling, and resource usage \cite{zhang2026towards,huang2025generalized,jiang2026intention}. Existing studies have investigated intent recognition, intent-aware information processing, and strategy adaptation, but these directions are developed relatively independently. A more integrated system framework is still needed to connect intent modeling, semantic evaluation, information updating, and strategy adaptation across heterogeneous services and changing contexts.

\item 
Intelligent complex systems integrate a highly coupled heterogeneous resource pool encompassing communication, computing, network, storage, and sensing \cite{saad2019vision}.  This integration maximizes system utility through mathematical-model optimization \cite{zhang2025resource,hu2024exploiting} and AI-driven learning optimization \cite{xu2025heterogeneous,lv2025resource, xu2025heterogeneous}. Furthermore, heterogeneous resource integration extends into the lifecycle of self-configuration and self-optimization. This lifecycle is driven by three methods: edge SLM automation \cite{lwin2026performance, lira2026network}, cognitive loop evolution \cite{zhang2026towards} and decentralized agentic collaboration \cite{xiao2026sanet}. Existing studies have explored heterogeneous resource integration and autonomous lifecycle management. However, a more integrated system framework is still needed to connect mathematical and AI-driven resource orchestration, cognitive loop evolution, and decentralized agentic collaboration across heterogeneous services and changing contexts.

\item 

Generative AI supports intelligent complex systems by learning system state distributions, reconstructing incomplete observations under uncertainty, and supporting semantic representations \cite{zhou2026digtwinai,gunduz2022beyond}. Agentic AI further extends these modeling capabilities toward intent deployment and closed-loop orchestration by integrating planning, tool use, and feedback-driven adaptation, while specialized agent organization and semantic-native coordination enable distributed collaboration\cite{zhou2026digtwinai,zhang2026agenticsurvey, leivadeas2022survey}. The major challenges in current research lie in trustworthiness, explainability, scalability, tool-use safety, and coordination reliability. These challenges point to unresolved issues in generated-state validation, constrained tool invocation, traceable agentic decision processes, and safe action execution.

\item 
Embodied AI grounds network intelligence in physical environments through closed-loop perception, decision, and action, supported by endogenous world models and task-critical semantic scheduling \cite{duan2022survey, ha2018world, gunduz2022beyond}. It transforms intelligent complex systems from passive data networks into active physical agents that can probe, model, and act within real-world constraints. Symbodied AI further elevates this capability by embedding human cognition, intent, and feedback into the decision loop through human-centric symbiotic alignment and intent-aware cognitive orchestration \cite{driess2023palm, sun2025collabvla, leivadeas2022survey}. This evolution shifts the focus from autonomous machine execution to human-machine co-evolution, where network resources and reasoning are organized around shared intent. However, current research still lacks unified frameworks for aligning human values with machine actions, ensuring intent safety, and orchestrating high-dimensional cognitive semantics over resource-constrained networks. These gaps point to the need for cognition-aware semantic scheduling, value-aligned learning, and trustworthy human-machine collaboration in symbiotic intelligent complex systems.

\end{itemize}

\section{Applications and Services Enabled by Intellicise Network-enhanced Intelligent Complex Systems}
\label{section5}

This section presents representative applications and services enabled by intellicise network-enhanced intelligent complex systems, which illustrates how the proposed architecture and key enabling technologies support intelligent operation, autonomous collaboration, and adaptive service delivery across diverse scenarios.

\subsection{Integrated Space-air-ground-sea Networks}
SAGSINs enhances the coverage and service continuity of future wireless networks by integrating heterogeneous and complex satellite networks, aerial networks, terrestrial networks, and maritime networks, thereby enabling ubiquitous connectivity across different spaces, regions, and scenarios \cite{wang2025toward, 11535035}. SAGSINs are typical large-scale complex systems characterized by multi-layer heterogeneous architectures, dynamic topology evolution, cross-domain resource coordination, and highly time-varying wireless channels \cite{wang2025task}. It is usually deployed in challenging environments where conventional terrestrial communication systems are difficult to establish, such as forests, deserts, oceans, remote mountainous areas, and disaster-stricken regions. Due to the coexistence of various communication platforms and the complex interactions among numerous network entities, SAGSINs exhibit strong nonlinearity, uncertainty, and dynamic adaptability. 
For SAGSINs, the main requirement is to maintain service continuity when network topology, link availability, and channel conditions change across different domains. Through a perception-communication-computing-actuation integrated paradigm \cite{xu2026vtfsc}, the control plane coordinates satellite, aerial, terrestrial, and maritime resources according to current connectivity and service priorities. Heterogeneous resource integration provides a unified resource basis, while self-configuration and self-optimization adjust network connections and transmission strategies as the environment changes. Moreover, semantic extraction limits the information transmitted over constrained satellite and aerial links to that required by the current task, improving service reliability under dynamic conditions.


\subsection{Industrial Internet of Things}
The IIoT is a complex system composed of massive industrial equipment, sensors, controllers, robots, edge computing nodes, and cloud platforms \cite{he2025graph}. The system features large-scale data generation, heterogeneous device types, stringent real-time requirements, and high reliability demands. During production, equipment states, manufacturing tasks, and resource allocation strategies continuously change, while local failures may propagate throughout the entire system. Such characteristics make the Industrial Internet a representative cyber-physical complex system. 
In industrial environments, communication performance is directly tied to production control. Equipment states, fault information, and control commands must be processed within strict latency and reliability constraints. Intellicise networks use the computation plane to support local processing and fault diagnosis at edge nodes, and the control plane coordinates manufacturing tasks, equipment actions, and resource allocation. Agentic AI further uses operational feedback to update decisions and actions during production. Together with semantic extraction of key equipment states and abnormal features, these capabilities support rapid fault response and stable production line operation \cite{hu2021manufacturing}.

\subsection{Smart Healthcare}
Smart healthcare systems integrate smart textiles, wearable devices, physiological sensors, edge computing platforms, healthcare databases, and medical personnel into a highly interconnected service ecosystem. The interactions among human physiological processes, medical devices, diagnostic systems, and healthcare services create a complex system with significant heterogeneity and dynamic behavior. Continuous monitoring generates massive multimodal medical data, while healthcare applications require strict guarantees on communication reliability, security, and timeliness. Smart textiles themselves are sophisticated systems integrating sensing, computing, communication, and data processing capabilities. 
Physiological data (e.g., remote healthcare monitoring) are continuous and sensitive, but only part of them requires immediate attention from medical personnel. In this setting, intellicise networks use the data plane to organize information collected from wearable devices, environmental sensors, and healthcare records, while the security plane protects its storage, access, and exchange. Based on semantic KBs, semantic extraction identifies abnormal physiological states, key indicators, and diagnosis-related information for prioritized transmission \cite{xiao2023semantic}. This allows wearable and remote monitoring systems to provide timely medical information while limiting communication and energy consumption.


\subsection{Intelligent Transportation}
Intelligent transportation systems consist of vehicles, roadside units, traffic lights, base stations, edge servers, traffic management platforms, and various sensing devices that continuously exchange information and collaboratively make decisions. Owing to high vehicle mobility, rapidly changing network topology, uncertain traffic behaviors, and large-scale interactions among multiple entities, intelligent transportation systems are widely regarded as complex adaptive systems. Such systems must simultaneously process heterogeneous information sources, including vehicle states, road conditions, traffic events, and user demands, while responding to dynamic environmental changes in real time. 
Road safety depends on the timely delivery of rapidly changing information, including hazardous object locations, abnormal driving behaviors, and accident warnings. Intellicise networks address this time sensitivity through intent understanding, which determines the relevance and priority of information according to the current driving or traffic-management task. The control plane coordinates transmission and computing resources according to urgency, while self-optimization updates these strategies as vehicle density, network topology, and road conditions change \cite{chen2022review}. These capabilities support vehicle-road collaboration, accident-risk prediction, and real-time traffic control.

\subsection{Digital Twin}
A digital twin establishes a real-time synchronized relationship between physical entities and their virtual counterparts, creating a complex system consisting of physical devices, sensing networks, data platforms, simulation models, and decision-making mechanisms. The system continuously collects, processes, and synchronizes massive amounts of sensing, control, operational, and interaction data from multiple sources and hierarchical levels. As the scale of the digital twin expands, maintaining consistency between physical and virtual spaces becomes increasingly challenging due to growing communication overhead and data redundancy. Therefore, digital twins represent a highly interconnected and data-intensive complex system. Intellicise networks leverage semantic compression and knowledge-based reasoning techniques to identify and synchronize only the key semantic information that affects twin-state evolution and operational decisions \cite{boje2020semantic}. Compared with transmitting all raw data, this approach substantially reduces communication costs, improves data utilization efficiency, and enhances the responsiveness of digital twin systems. Furthermore, intellicise networks provide a holistic and scalable semantic framework that dynamically considers data value and semantic relationships across device, network, service, and decision-making layers.

\subsection{Embodied Agent Communications}
Embodied intelligence systems consist of intelligent agents, robotic platforms, sensors, actuators, edge computing resources, and surrounding environments that interact in closed feedback loops. These systems continuously perform perception, cognition, decision-making, and action execution while processing multimodal information such as vision, speech, touch, position, and motion. The strong coupling among sensing, communication, computation, and control processes makes embodied intelligence a typical complex system. The complexity further increases in scenarios involving multi-robot collaboration, swarm intelligence, and autonomous systems, where large numbers of agents must coordinate and exchange information efficiently. One of the major challenges is the transmission of high-dimensional and heterogeneous multimodal data, which places considerable pressure on communication bandwidth and latency. 
Embodied agents place communication directly within the physical control loop, where delayed or irrelevant information can affect the resulting action. Intellicise networks use the intelligence plane to support multimodal semantic representation and task reasoning, while the computation plane coordinates local and edge inference under control-latency constraints. Cross-modal feature fusion and attention mechanisms transform vision, speech, touch, position, and motion information into compact representations for planning and control \cite{xu2026vtfsc}. Through embodied AI, intellicise networks connect the exchanged semantic information with physical execution, allowing communication performance to be assessed through task completion.


\section{Case Study: Intellicise Networks for Embodied Agent Communications}
\label{section6}

\begin{figure}
\centering
\includegraphics[width=\linewidth,height=.58\textheight,keepaspectratio]{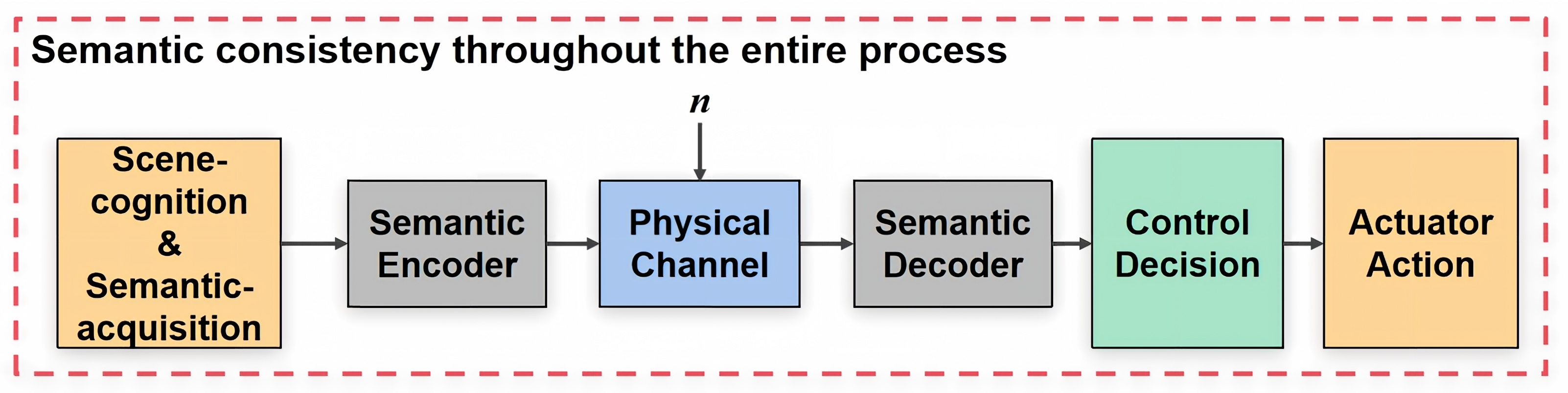}
\caption{Joint sensing--computing--communication--control (JSCCC) for embodied agent communications. The sensing module acquires RGB observations and extracts task-relevant semantic features. The communication module conveys these features over the physical channel. The computing module estimates the three-dimensional target position from the received features, and the control adapter converts this position into joint commands through inverse kinematics. The shared semantic interface connects information delivery, task-oriented inference, and physical action while bypassing image reconstruction and receiver-side image-based object detection.}
\label{fig_JSCCC}
\end{figure}

Embodied intelligence systems that treat communication separately from perception and control may experience a mismatch between information delivery and control requirements.
Yet communication is indispensable in tasks such as multi-agent networking and cloud-based decision transmission.
Existing communication networks are hindered by transmission efficiency and latency, which compromise real-time control performance and ultimately reduce task success rates.
Against this background, we present an embodied-agent case study to illustrate how task-relevant semantic representations connect sensing, computing, communication, and control within the architecture introduced in Section \ref{section3}. The case study focuses on the interaction between information delivery and task execution, providing a concrete example of the proposed architectural principles.

\subsection{Presented Scheme for Embodied Agent Communications}

As illustrated in Fig. \ref{fig_JSCCC}, we present a Joint Sensing-Computing-Communication-Control (JSCCC) scheme that integrates sensing, computing, communication, and control through a shared semantic representation. The received features support task-oriented inference, whose output is used to generate control commands. The usefulness of the transmitted representation is therefore assessed through task execution rather than image reconstruction alone. The scheme comprises the following four functional components:

\begin{itemize}
    \item \textit{Sensing:} The sensing and semantic-acquisition module captures RGB observations of the scene and extracts task-relevant visual features, forming a semantic representation for subsequent communication, computation, and control.

    \item \textit{Computing:} The computing module performs task-oriented inference on the received semantic features to obtain the information required for control. In the evaluated implementation, a multilayer perceptron estimates the three-dimensional target position directly from these features, providing the control adapter with the spatial information needed for action generation.

    \item \textit{Communication:} The semantic representation is encoded and transmitted over the physical channel. The received semantic features are then provided to the computing module for task-oriented inference, without reconstructing the original sensory input. This design connects information delivery to the requirements of the downstream task.

    \item \textit{Control:} The control adapter converts the target position estimated by the computing module into joint commands through inverse kinematics. These commands drive the robotic actuator to perform the reaching task. This interface links task-oriented inference to physical action without requiring receiver-side image reconstruction or image-based object detection.
\end{itemize}

In the evaluated implementation, the sensory input is an RGB image, and the semantic representation is extracted by a SwinJSCC encoder. At the receiver, a multilayer perceptron estimates the three-dimensional target position from the received features. This position is then converted into joint commands through inverse kinematics, without reconstructing the image or performing receiver-side image-based target detection.

From the perspective of Section \ref{section3}, semantic extraction corresponds to the perception and cognition layer, while semantic transmission corresponds to the access and transmission layer. Receiver-side inference provides a functional instance of the interconnection and computing layer, and the conversion of task-relevant information into executable actions corresponds to the decision and control layer. These operations draw on the intelligence plane for learned representations and inference, the computation plane for processing, and the user plane for carrying task-related information. Observations and execution states constitute the data flow, while the task objective and execution outcome connect information processing to the task flow. Thus, JSCCC illustrates how a shared semantic representation links functions across layers and planes according to downstream task requirements.

\subsection{Simulation Results and Analysis}

We evaluate the schemes on a simulated reaching task using a six-degree-of-freedom UR5 robot in PyBullet. An RGB image provides the visual observation, and the semantic communication path is evaluated over an additive white Gaussian noise (AWGN) channel. Three schemes are compared: the presented JSCCC scheme, a traditional pipeline comprising JPEG compression, LDPC coding and decoding, image reconstruction, and visual geometric reaching (VGR), and a semantic communication pipeline comprising SwinJSCC image reconstruction followed by the same VGR controller.

For the semantic schemes, the representation precision is varied over $q \in \{1,2,4,8,16,32\}$ bits per feature element. The evaluated latent representation contains $1024 \times 128$ elements, corresponding to a nominal representation size of $q \times 1024 \times 128$ bits per observation. JPEG quality is adjusted in the traditional pipeline according to a budget derived from this nominal representation size. The communication resource setting is therefore characterized by the nominal representation size per observation.

The reaching success rate is defined as the proportion of trials in which the end effector approaches within $5$~cm of the estimated desired position during the episode. End-to-end episode runtime is measured from observation acquisition to episode termination, covering communication-module processing, receiver-side inference, and simulated action execution. The reported runtime is averaged over the evaluated trials.

\begin{figure}
\centering
\includegraphics[width=\linewidth,height=.58\textheight,keepaspectratio]{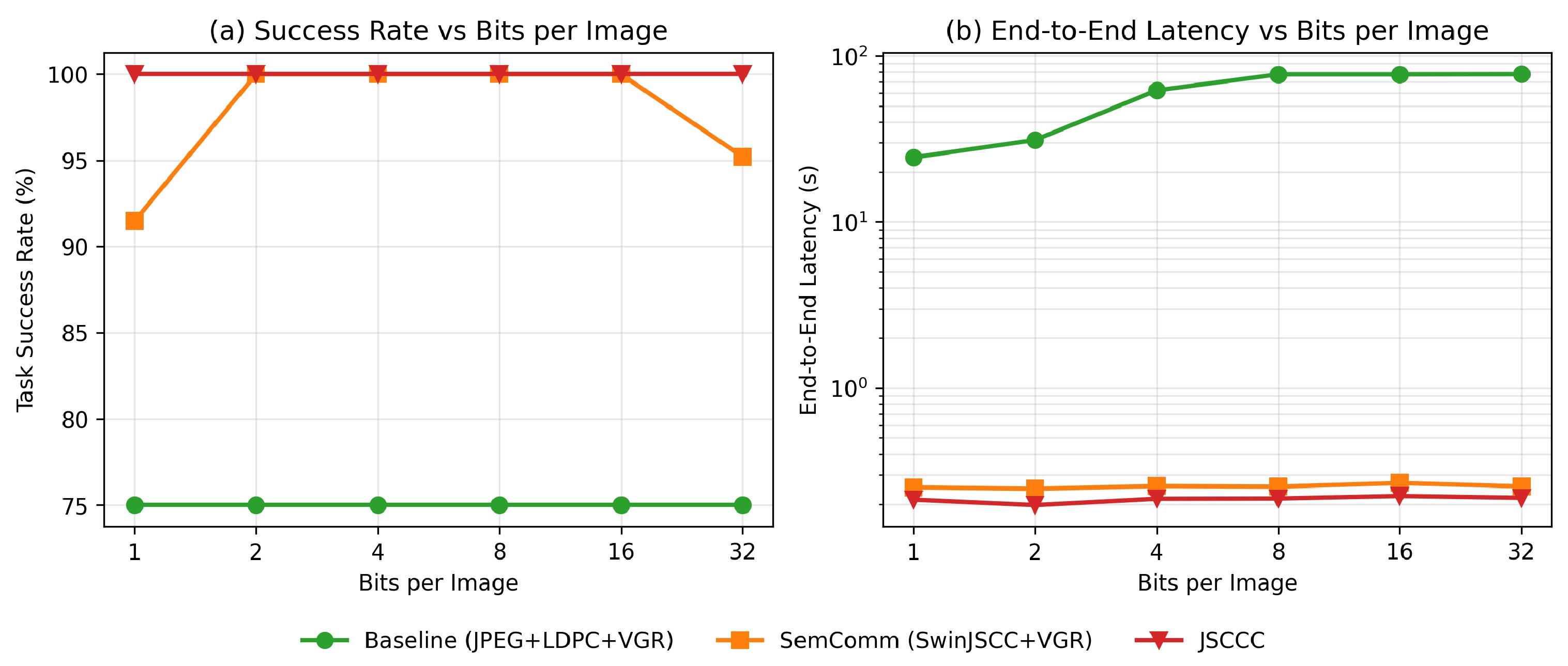}
\caption{Comparison of reaching success rate and end-to-end episode runtime for JSCCC and two cascaded communication--control pipelines under the tested representation settings. The reference pipelines use JPEG--LDPC or SwinJSCC image reconstruction followed by image-based target detection and VGR control, whereas JSCCC predicts the target position from received features without image reconstruction. The comparison examines how the interface between communication and control affects reaching performance and processing requirements.}
\label{fig_simulation_result}
\end{figure}

Fig. \ref{fig_simulation_result} compares reaching success rate and end-to-end episode runtime across the tested representation settings. The success-rate comparison examines whether the transmitted representation preserves the information required for reaching, while the runtime comparison shows the processing cost of the complete sensing--communication--computing--control pipeline. In JSCCC, the received features support target-position prediction directly, bypassing image reconstruction and receiver-side image-based detection and simplifying the processing path from information reception to action generation.

This case study illustrates how a task-relevant semantic representation serves as a shared interface linking sensing, computing, communication, and control. The task objective guides the information required by the receiver, and the received representation supports action generation through the control adapter. This connection reflects the semantic-native and task-oriented design ideas of the proposed architecture, linking cross-layer information processing to task execution.



\section{Future Research Directions}
\label{section7}
Several issues remain to be addressed for the further development of intellicise network-enhanced intelligent complex systems. In this section, future research directions concerning compatibility with traditional communication networks, generalization across scenarios, interpretability of AI models, and security and privacy are discussed as follows.
\subsection{Compatibility between Traditional Communication Networks and Intellicise Networks}
The deployment of intellicise networks will not be an instantaneous replacement of existing communication infrastructure. Instead, it must coexist with and gradually evolve from the current generation of networks. This coexistence introduces significant compatibility challenges across multiple dimensions. At the protocol level, traditional networks rely on strict layer isolation and bit-accurate transmission, whereas intellicise networks emphasize cross-layer semantic interaction and task-oriented information exchange. Bridging this fundamental architectural gap requires the design of adaptive protocol translation mechanisms that enable legacy devices to participate in semantic-aware communication without requiring hardware upgrades. At the resource management level, existing networks allocate spectrum, computing, and storage resources independently, while intellicise networks couple these heterogeneous resources dynamically through semantic KBs and task-driven scheduling. Ensuring backward compatibility in resource abstraction and orchestration is therefore critical for smooth migration. Furthermore, the service continuity of existing applications must be preserved during the transition, which demands that intellicise networks support both semantic transmission and traditional bit-level transmission modes simultaneously \cite{alliance2025network}.

\subsection{Generalization of Intelligent Complex Systems for Various Scenarios}
The proposed intelligent complex system framework is designed to be universally applicable across diverse scenarios including integrated space-air-ground-sea networks, industrial Internet of Things, smart healthcare, intelligent transportation, digital twins, and embodied agent communications. However, each scenario presents distinct physical characteristics, service requirements, and environmental constraints. A framework instance that performs well in one scenario may not transfer effectively to another without substantial retuning. This generalization challenge stems from the distribution mismatch between training and deployment environments, which is further compounded by the variability of wireless channel conditions, user behavior patterns, and task semantics across domains. Recent advances in distributionally robust optimization and causality invariant learning have shown promise for improving model robustness against semantic drift and channel variability \cite{le2026distributionally, nguyen2025knowledge}. Transfer learning and meta-learning frameworks also enable models to adapt to new tasks or environments using minimal local data, which reduces the overhead of scenario specific customization \cite{zheng2025semantic}. Future research should investigate adaptive framework instantiation mechanisms that can automatically adjust the emphasis of each functional plane and information flow structure based on the specific characteristics of the target scenario, thereby moving from a one size fits all architecture toward a truly scalable and scenario aware intelligent system.

\subsection{Interpretability of AI Models in Intelligent Complex Systems}
The intelligence of future intelligent complex systems depends heavily on AI models such as large language models, vision language models, and multi agent frameworks. However, the inherent opacity of these models poses a fundamental challenge to the trustworthiness and operational reliability of the entire system. In intelligent complex systems, decision making spans multiple layers from perception and cognition through to decision and control, and a single misaligned decision at any layer may cascade into system level failures. Without interpretable decision processes, operators cannot diagnose the root cause of abnormal behaviors, validate the rationality of AI actions, or comply with emerging regulatory requirements for AI transparency. Recent research on explainable AI in communication networks has begun to address this gap by developing frameworks that provide human understandable explanations for AI decisions in real time \cite{sun2025advancing, fiandrino2022toward}. Nevertheless, applying these techniques to the cross domain, multi layer, and tightly coupled architecture of intellicise networks remains an open problem. Future research should focus on building knowledge driven interpretable reasoning frameworks that can trace semantic information flows across layers, attribute decisions to specific KBs and model components, and generate actionable explanations for network operators. By manifold learning for intelligent complex systems, mapping of semantic information flow to topology of models or convergence of manifold based on KBs of complex systems could be achieved \cite{xie2026bridging,xie2025mhc}. Such interpretability is not merely a desirable property but a prerequisite for achieving self configuration and self optimization in intelligent complex systems.

\subsection{Security and Privacy of Intelligent Complex Systems}
The expansion of intelligence in future networks also enlarges the attack surface that must be defended. In intelligent complex systems, security threats extend far beyond the physical layer and link layer to encompass data integrity, model integrity, KB integrity, and the authenticity of cross plane interactions \cite{xiao2026rf}. The introduction of large language models and multi agent frameworks creates new vulnerabilities such as prompt injection attacks, adversarial manipulation of semantic KBs, and model extraction from edge devices \cite{guo2024survey, joshi2026securing}. More critically, agentic eavesdroppers that combine planning, memory, external knowledge, semantic reasoning, and generative priors can adaptively infer private information from intercepted signals\cite{tang2026eavesdroppers},
Furthermore, the semantic nature of information exchange in intellicise networks means that traditional cryptographic protection mechanisms designed for bit level transmission are insufficient to guard against semantic level eavesdropping and KB leakage. Cross-layer security frameworks that jointly optimize physical layer confidentiality and application layer semantic protection have begun to address this gap \cite{meng2023physical,naqvi2026security}. Privacy preserving mechanisms such as federated learning, differential privacy, and homomorphic encryption are also essential for protecting sensitive information in scenarios involving personal health data, location data, or behavioral patterns. Future research should develop security architectures that are co-designed with the intelligent plane and the semantic KB, enabling proactive threat detection, adaptive defense, and trust assessment across all planes and layers of the intelligent complex system. In particular, the researches of LLM-enhanced attacks and LLM-based defenses should be systematically investigated to ensure that security mechanisms can anticipate and counter increasingly autonomous and adaptive adversaries.

\section{Conclusions}
\label{section8}
In this article, we have presented a systematic perspective on future intelligent complex systems from the viewpoint of intellicise networks. We first proposed a cross-domain intelligent communication network architecture that provides a structured foundation for understanding and designing intellicise network-enhanced intelligent complex systems. Grounded in information theory, system theory, game theory, and cybernetics, the proposed architecture integrates a four-layer organizational framework, six multi-functional planes, and four novel information flows. Building upon this architecture, we reviewed key technological evolutions and discussed representative applications and services to illustrate the potential of intellicise networks in supporting diverse complex scenarios. Specifically, the architecture organizes the required system functions, the enabling technologies translate them into practical capabilities, and the representative applications show how these capabilities support different complex scenarios.
We further presented a case study to show the effectiveness of intellicise networks in improving the performance of embodied agent communications. Despite the significant progress in this emerging field, several fundamental issues remain open. Finally, we summarized future research directions concerning compatibility with traditional networks, cross-scenario generalization, AI model interpretability, and system security and privacy.

\bibliographystyle{unsrtnat}
\bibliography{ref}
\end{document}